\documentclass{article}

\usepackage{arxiv}

\usepackage{amsmath}
\usepackage[utf8]{inputenc} 
\usepackage[T1]{fontenc}    
\usepackage{hyperref}       
\usepackage{url}            
\usepackage{booktabs}       
\usepackage{amsfonts}       
\usepackage{nicefrac}       
\usepackage{microtype}      
\usepackage{lipsum}		
\usepackage{graphicx}

\usepackage{algorithm} 
\usepackage{algorithmicx}
\usepackage{algpseudocode}
\usepackage{tabularx,array,booktabs}
\usepackage{tabstackengine}
\usepackage{subcaption}
\usepackage{caption}
\usepackage{graphicx}
\usepackage{float}
\usepackage{longtable}
\usepackage{acronym}
\usepackage[capitalise]{cleveref}
\usepackage{overpic}
\usepackage{colortbl}
\usepackage{cellspace}
\usepackage{cancel}
\usepackage{xcolor}
\usepackage{authblk}
\usepackage{multirow}

\title{Lagrangian-Eulerian Multi-Density Topology Optimization with the Material Point Method}

\date{} 					

\author[*,1, 4]{Yue Li}
\author[*,1, 2]{Xuan Li}
\author[*,1, 2]{Minchen Li}
\author[2]{Yixin Zhu}
\author[3]{Bo Zhu}
\author[1, 2]{Chenfanfu Jiang}

\affil[*]{Equal contributors}
\affil[1]{University of Pennsylvania}
\affil[2]{University of California, Los Angeles}
\affil[3]{Dartmouth College}
\affil[4]{ETH Zürich}

\newcommand{\LETO}{LETO }
\newcommand{\LETON}{LETO}

\usepackage{xspace}
\makeatletter
\DeclareRobustCommand\onedot{\futurelet\@let@token\@onedot}
\def\@onedot{\ifx\@let@token.\else.\null\fi\xspace}
\def\eg{\emph{e.g}\onedot} 
\def\ie{\emph{i.e}\onedot} 
 
\def\etc{\emph{etc}\onedot} 
\def\wrt{w.r.t\onedot} 
\def\etal{\emph{et al}\onedot}
\makeatother

\begin{document}
\maketitle

\begin{abstract}
In this paper, a hybrid Lagrangian-Eulerian topology optimization (\LETON) method is proposed to solve the elastic force equilibrium with the Material Point Method (MPM).
\LETO transfers density information from freely movable Lagrangian carrier particles to a fixed set of Eulerian quadrature points. This transfer is based on a smooth radial kernel involved in the compliance objective to avoid the artificial checkerboard pattern.
The quadrature points act as MPM particles embedded in a lower-resolution grid and enable a sub-cell multi-density resolution of intricate structures with a reduced computational cost.
A quadrature-level connectivity graph-based method is adopted to avoid the artificial checkerboard issues commonly existing in multi-resolution topology optimization methods.
Numerical experiments are provided to demonstrate the efficacy of the proposed approach.
\end{abstract}


\section{Introduction}

Topology optimization is experiencing a rapid advance over the past few years, thanks to the collision of waves between next-generation computing infrastructure and high-performance simulation software. A surge of recent work has been creating various computing infrastructures capable of accommodating topology optimization applications with a super-scale resolution---millions to one billion of material voxels---on parallelizable data structures (\eg, \cite{aage2017giga,liu2018narrow,wu2015system}). These density-based approaches stem from the conventional Solid Isotropic Material with Penalization Method (SIMP) \cite{sigmund200199,andreassen2011efficient} and naturally fall into Eulerian methods, owing to their geometric representation of the material evolution on a fixed grid. Level set\cite{osher1988fronts}-based methods\cite{wang2003level,allaire2004structural,luo2008level,van2013level} are also Eulerian due to an implicit representation of the topology on grid nodes. On the other hand, Lagrangian geometries are increasingly attracting attention. For example, particles (\eg, in smoothed-particle hydrodynamics (SPH) \cite{gingold1977smoothed}) can explicitly track the structural evolution under the guidance of material derivatives. Tracking explicit meshes is also a promising direction thanks to the advent of high-performance meshing software \cite{christiansen2014topology}.
\subsection{Hybrid Representation}
Despite extensive research, Eulerian approaches have limited capability in capturing intricate structures, especially when the problem requires fine features that are hierarchical, codimensional, and can emerge from a nihil. On the other hand, Lagrangian representations suffer from a lower computational performance. Analogous to their computational physics counterparts (\eg, in computational fluid dynamics), Eulerian approaches are not naturally adaptive to subgrid features, whereas Lagrangian methods face challenges in establishing differential stencils that are geometrically symmetric and numerically accurate.
\par
In computational physics, researchers face the same dilemma regarding the choice of data structures and the corresponding numerical stencils when simulating large-scale fluids and solids. 
This dilemma further triggered the invention of a bank of hybrid Lagrangian-Eulerian methods, such as Particle-In-Cell (PIC) / Fluid-Implicit-Particle (FLIP) methods \cite{harlow1962particle,brackbill1988flip} and Material Point Methods (MPM) \cite{sulsky1995application,de2019material}, which are featured by an Eulerian background grid as a scratch pad and a set of Lagrangian particles to track geometry and topology. 
By conducting data transfers between the two representations, a hybrid Lagrangian-Eulerian scheme can typically leverage both sides' merits, enabling flexible and robust numerical solutions \cite{brackbill1988flip,sulsky1995application,zhang2016material}.
\par
Motivated by such a design philosophy, some hybrid methods are also proposed for topology optimization.
For example, the Moving Morphable Component (MMC) method \cite{guo2014doing, zhang2016lagrangian,zhang2017explicit,zhang2017multi,zhang2017new,lei2019machine} aims to substantially reduce the number of design variables by optimizing component-wise distributions. It represents structures by unions of superellipse level sets -- low dimensional morphable components that can move, deforming, and overlapping to track topology changes. The explicit geometric information also helps control the minimum length scale \cite{zhang2016minscale}.
MMC can produce results with sharp features with attractive convergence and timing profiles. 
However, to acquire sophisticated geometry features, a large number of components, \ie, design variables, is necessary. 
On the other hand, the Moving Node Approach (MNA) \cite{overvelde2012moving} represents the target shape with a set of mass nodes, of which the positions are optimized to search for an optimal structure. In MNA, the quadrature is a set of regularly sampled discretization nodes, on which the densities are computed according to the clustering of mass nodes. 
However, such an approach can lead to results with many isolated mass nodes disconnected from the main structure, which can only be cleaned up using an extra post-processing step. 
\par
Inspired by these hybrid methods, a new hybrid Lagrangian-Eulerian topology optimization method---\LETO is proposed in this paper. This new approach optimizes material distributions over a design domain by evolving a set of material carrier particles on a background Cartesian grid. 
With MPM applied for solving the static equilibrium,
another set of particles, each carrying a temporally varying density, is evolved as a Lagrangian representation of material distribution, eliminating redundant, isolated particle blobs.
The Lagrangian-Eulerian nature of the framework enables the communication between the moving particles and the fixed background MPM quadrature points by transferring the density values through interpolation functions. 
More specifically, as the carrier particles move and change their densities, the quadrature points' density values are updated accordingly, naturally providing sub-cell density resolution.
As shown in the experiments, \LETO tends to generate structures with rich branching fibers with low compliance.

\subsection{Topology Optimization with the Material Point Method}
When applying the adjoint method on topology optimization for sensitivity analysis, elasticity simulation is required in each iteration to obtain the nodal displacements under force equilibrium given a material distribution and an external load. A static elasticity solver can be applied since inertia effects are often ignored. While traditional topology optimization methods often use a grid-based Finite Element discretization, we adopt MPM for the static setting with the sub-cell resolution achieved by assigning a different density value to each quadrature particle. The static force equilibrium is further solved with a variational formulation that guarantees robustness and stability. 

\paragraph{Spatial Discretization}
Traditional topology optimization methods, including both density-based approaches \cite{sigmund200199,buhl2000stiffness} and level set-based approaches \cite{guo2014doing,wang2003level}, often apply grid-based Finite Element discretization for the static solve. In FEM, the same material density is assigned for all quadratures in every single cell. Therefore, the domain boundaries are formed with jagged finite element edges, and even plotting zero-level contour still results in jagged boundaries \cite{maute2013topology}. A grid with higher resolution is often required to alleviate these artifacts, which increases the computational cost.

MPM is a hybrid Lagrangian-Eulerian method widely used in different fields, \eg, computer graphics \cite{stomakhin2013material,wolper2019cd}, civil engineering \cite{abe2014material,zabala2011progressive}, mechanical engineering \cite{sulsky1995application,guilkey:2003:implicit-mpm,sinaie2018validation,chen2002evaluation}. With the capability of handling large deformation\ \cite{nair2012implicit,charlton2017igimp,nguyen2016voronoi,sadeghirad2011convected,soga2016trends,lian2012adaptive}, topology changes, and coupled materials, MPM has been considered as one of the top choices in various physics-based simulations, including fracture \cite{guo2006three,wolper2019cd,yang2014improved,long2019using,long2016representing,homel2017field}, viscoelastic and elastoplastic solids \cite{fang2019silly,burghardt2012nonlocal}, incompressible materials\ \cite{kularathna2017implicit,zhang2017incompressible}, high explosive explosion\ \cite{ma2009simulation}, snow \cite{stomakhin2013material,gaume2018dynamic,gaume2019investigating}, granular material \cite{bardenhagen2000material,klar2016drucker,yerro2019runout,zhang2009material} and mixtures \cite{gao2018animating,tampubolon2017multi,bandara2015coupling}.
In MPM, Lagrangian particles, which are also known as material points, are used to track quantities like mass, momentum, and deformation. On the other hand, a regular Eulerian grid is built to evaluate force and update velocity at each time step. Particle quantities are then updated from the interpolation of nodal quantities.
MPM's convergence was demonstrated computationally and explained theoretically with a smooth, e.g., quadratic B-spline, basis for grid solutions\ \cite{steffen2008analysis}, which was further verified with manufactured solutions\ \cite{wallstedt2009order}.
MPM is applied as the spatial discretization in this work. We also describe a static formulation for directly solving the force equilibrium, which allows defining quadrature-wise density per cell to take advantage of the sub-cell resolution. As shown in numerical experiments, \LETO achieves a comparable convergence speed with lower structural compliance.

\paragraph{Optimization and Nonlinear Integrators}
Another long-standing challenge of topology optimization is to optimize structures undergoing large deformations, requiring a nonlinear elasticity model and the nonlinear equilibrium constraints.
This has become increasingly meaningful with the increasing need for material and structural design in soft robotics\cite{scharff2019robot}, wearable devices, and even space antennas, \etc.
With large nonlinear deformations, the force equilibrium is more challenging to solve as it often leads to numerical instabilities, and the optimization itself will converge slower. 

Numerical integration of partial differential systems can often be reformulated variationally into an optimization problem. These methods can often achieve improved robustness, accuracy, and performance by taking advantage of well-established optimization approaches. 
Simulation methods are increasingly applying this strategy to simulate both fluid \cite{batty2007fast}, and solid \cite{gast2015optimization} dynamics, which often enable large time step sizes. The static solve simply corresponds to infinitely large time step size in a dynamic time-stepping point-of-view. Our method also takes advantage of optimization integrators to solve for the static equilibrium to high accuracy robustly. 

For nonlinear optimization problems, Newton-type methods are often used because they can deliver quadratic convergence when the intermediate solution becomes close to the local optima. However, when the initial configuration is far from a local optimum, which is often true in static solves, Newton's method may fail to provide a proper search direction as the Hessian can be indefinite \cite{wang2019hierarchical,liu2017quasi,liu2017quasi,smith2018stable}. Teran \etal \cite{teran2005robust} proposed a positive definite fix to project the Hessian to a symmetric positive definite form to guarantee that a descent direction can be found. This method is referred to as projected Newton (PN) throughout the paper and is applied in the static solve. 

\subsection{Artificially Stiff Patterns}

Traditional density-based topology optimization methods considering one degree of freedom for density per element may form checkerboard density patterns across elements. 
Such solutions are indeed optimal mathematically but meaningless in reality. To avoid the checkerboard pattern, filters \cite{sigmund200199,andreassen2011efficient} were proposed to smooth the density or gradient field. 
These filters can be applied independent of the sensitivity analysis or encoded into the objective \cite{wu2015system}, where the latter one defines a consistent optimization problem.
In this paper, the SPH kernel-based density transfer between carrier particles and quadrature points serves the same purpose and is explicitly encoded in the objective.
\par
\begin{figure}
    \centering
    \includegraphics[width=0.5\textwidth]{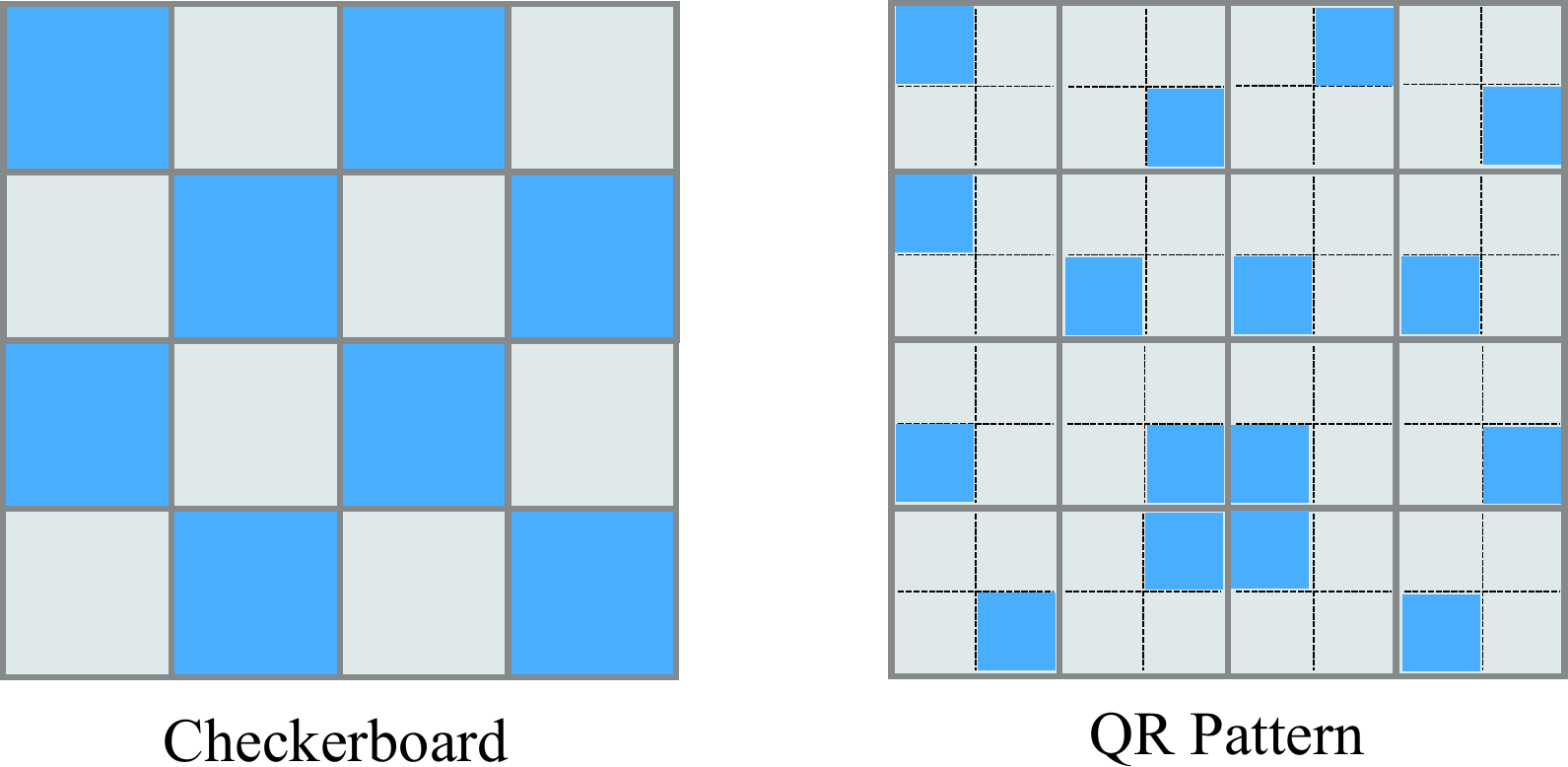}
    \caption{An illustration of the checkerboard issue and the QR pattern issue. In both cases, the density fields are continuous from the view of the grids}
    \label{fig:checker_vs_qr}
\end{figure}

When multiple densities are modeled inside one element to generate higher-resolution details with a relatively low computational cost \cite{Nguyen2009}, sub-cell-level checkerboard issues, also called the QR patterns \cite{Gupta2018}, may appear.
Such artifacts happen on a sub-cell level where disconnected interfaces between material components may form inside one element at the quadrature level.
However, from the simulation grid view, the disconnected components are connected, producing inaccurate compliance measurement at static equilibrium, which can lead to results with large compliance when tested in practice or simulated with higher resolution. An illustration of the checkerboard issue and the QR pattern issue is shown in \cref{fig:checker_vs_qr}.
Existing solutions for avoiding QR patterns include increasing the filter radius of gradient filters \cite{Nguyen2009}, changing penalty power in SIMP formulation \cite{Gupta2018} and using higher-order elements \cite{Groen2016}.
If applied naively, our method may encounter this kind of artifact as well if no treatment is applied since the resolution of MPM quadratures is higher than the background grid. To take advantage of the multi-resolution nature of MPM while keeping the formulation simple, a graph-based narrow-band filter and a connectivity correction algorithm are developed: the set of material points is considered as a graph and only one major connected component is preserved during optimization. A secondary correction is applied after the optimization.

\subsection{Summary}
In summary, a novel hybrid Lagrangian-Eulerian framework \LETO is proposed for compliance-based topology optimization with MPM.
The hybrid representation of the material density field enables both the flexibility of Lagrangian models and Eulerian methods' computational efficiency.
The MPM discretization introduces a multi-density scheme naturally, provides a unified treatment for both linear and nonlinear topology optimization, and supports optimizing fully nonlinear compliance, enabling robust and accurate optimization of structures undergoing large deformations.
A graph-based narrow-band filter and a connectivity correction algorithm are applied during and after the optimization to eliminate QR-patterns.
The numerical experiments show that the resulting scheme can generate sub-cell structures with mechanical performances that sometimes rival conventional methods at a comparable computational cost. 

\section{problem statement and method overview}

\subsection{Problem Statement} 

The general objective of compliance-based topology optimization is to seek for a material distribution $\mathbf{\rho}$, a scalar field representing the material density at each point on a design domain $\Omega$, to obtain the minimal structural compliance $c(\mathbf{\rho}, \mathbf{u})$, or equivalently, the least strain energy $e(\mathbf{\rho}, \mathbf{u})$, under force equilibrium between external force load $\mathbf{f}$ and internal elasticity force $-\frac{\partial e}{\partial \mathbf{u}}$ with displacement $\mathbf{u}$:
\begin{equation}
    \underset{\mathbf{\rho}}{\text{min}}\:\:c(\mathbf{\rho}, \mathbf{u}) = e(\mathbf{\rho}, \mathbf{u})
    \quad \text{s.t.} \quad
    \begin{cases}
        \frac{\partial e}{\partial \mathbf{u}}(\mathbf{\rho}, \mathbf{u}) = \mathbf{f} \\
        \mathbf{Du} = \mathbf{0}\\
        V(\mathbf{\rho}) \le \hat{V}.
    \end{cases}
    \label{eq:formulation}
\end{equation}
Here $\mathbf{Du} = \mathbf{0}$ is the discretized Dirichlet boundary condition where $\mathbf{D}$ selects the zero displacement nodes, $V(\mathbf{\rho}) = \int_{\Omega_0} \rho d\mathbf{X}$ is the total volume of the structure, and $\hat{V}$ is an upper bound specified by users to avoid trivial solutions\cite{sigmund200199}.
Usually, $\rho$ is expected to be close to either $0$ or $1$ for manufacturing, which potentially makes the problem non-smooth. The density field $\rho$ can be discretized and further parameterized by any set of design variables. 
For example, the traditional SIMP method assumes that each finite element has a uniform density and directly uses cell densities as design variables. In this paper, a set of movable Lagrangian particles carried with density sources is used as design variables.

The strain energy of the material under a displacement field $u$ is defined as
\begin{equation}
    e(\mathbf{\rho}, \mathbf{u}) = \int_{\Omega} \Psi(\mathbf{F}) d\mathbf{X},
\end{equation}
where $\Psi$ is the elastic energy density determined by the underlying constitutive model, and $\mathbf{F}$ is the deformation gradient defined as
\begin{equation}
    \mathbf{F} = \frac{\partial \mathbf{x}}{\partial \mathbf{X}} = \mathbf{I} + \frac{\partial \mathbf{u}}{\partial \mathbf{X}}
\end{equation}
through the world and material space coordinates $\mathbf{x}$ and $\mathbf{X}$ with $\mathbf{u}(\mathbf{X}) = \mathbf{x} - \mathbf{X}$ and $\mathbf{I}$ is the identity matrix.
For linear elasticity, 
\begin{equation}
   \Psi_L(\mathbf{F}) = \mu ||\epsilon(\mathbf{F})||^2 + \frac{\lambda}{2}\text{tr}(\epsilon(\mathbf{F}))^2,
\end{equation}
where $\epsilon(\mathbf{F}) = \frac{1}{2}(\mathbf{F} + \mathbf{F}^T) - \mathbf{I}$ is the small strain, and the  Lam\'e parameters $\mu$ and $\lambda$ linearly relate to the Young's modulus $E$. 
However, linear elasticity is only accurate under infinitesimal deformation since rotation is also penalized, and the nonlinear stress-strain curve is not well captured. A nonlinear constitutive model should be used to model large deformation.
In this paper, we adopt the neo-Hookean hyperelasticity model:
\begin{equation}
    \Psi_\text{NH}(\mathbf{F})=\frac{\mu}{2}(\text{tr}(\mathbf{F}^T\mathbf{F}) - d) - \mu\log{J}+\frac{\lambda}{2}(\log{J})^2,
    \label{eq:NH}
\end{equation}
where $J=\det{\mathbf{F}}$, and $d=2$ or $3$ is the dimension of the problem.
\par
The compliance objective depends on both the density $\rho$ and the displacement $u$, which is generally nonlinear even for linear elastic materials as well as the static equilibrium constraint. 
Therefore, the adjoint method \cite{giles2000introduction} is often applied to avoid solving the nonlinear Karush-Kuhn-Tucker (KKT) system as in equality constrained optimization \cite{nocedal2006numerical}. It takes $\mathbf{u}$ as a function of $\mathbf{x}$ and cancels out $\frac{\partial \mathbf{u}}{\partial \mathbf{x}}$ by considering the searching process to be conducted only on the force equilibrium constraint manifold. 
Given an intermediate state $\mathbf{\rho}$, the PDE constraint has to hold by solving the displacement field $\mathbf{u}$ at static equilibrium for each iteration.
For linearly elastic materials, finding the static equilibrium results in solving a linear system of equations, which is generally considered the topology optimization's bottleneck.
Therefore, obtaining intricate structural features by increasing resolution demands extra computational powers or carefully designed implementations. \cite{aage2017giga, liu2018narrow}.
It becomes even more challenging for nonlinear hyperelastic materials because a nonlinear system of equations needs to be solved at each optimization iteration, leading to numerical instabilities.

\subsection{Lagrangian-Eulerian Multi-Density Topology Optimization}
A hybrid Lagrangian-Eulerian approach is proposed to establish a versatile topology optimization framework that can accommodate different elastic models to address the aforementioned challenges. 
In particular, the elastic potential as the compliance objective is optimized for both linear and highly nonlinear (\eg, neo-Hookean) elastic materials. 
A set of carrier particles is adopted to represent the material distribution and evolution. Each particle is a moving material sample carrying the information of position $\mathbf{x}^c$, density $\mathbf{\rho}^c$, and supporting radius.
This modifies the general formulation in \cref{eq:formulation} from a pure Eulerian representation, which directly optimizes the density field $\mathbf{\rho}$, to a hybrid Lagrangian-Eulerian form, which jointly optimizes $\mathbf{x}^c$ and $\mathbf{\rho}^c$ that define the material distribution $\mathbf{\rho}(\mathbf{x}^c, \mathbf{\rho}^c)$ over the design domain:
\begin{equation}
    \underset{\mathbf{x}^c, \mathbf{\rho}^c}{\text{min}}\:\:c(\mathbf{\rho}(\mathbf{x}^c, \mathbf{\rho}^c), \mathbf{u}) = e(\mathbf{\rho}(\mathbf{x}^c, \mathbf{\rho}^c), \mathbf{u})
    \quad \text{s.t.} \quad \begin{cases}
        \frac{\partial e}{\partial \mathbf{u}}(\rho(\mathbf{x}^c, \mathbf{\rho}^c), \mathbf{u}) = \mathbf{f} \\
        \mathbf{Du} = \mathbf{0}\\
        V(\mathbf{\rho}(\mathbf{x}^c, \mathbf{\rho}^c)) \le \hat{V}.
    \end{cases}
    \label{eq:formulation_hybrid}
\end{equation}

Using MPM as the static equilibrium solver, the design domain is discretized with an Eulerian background grid and a set of uniformly sampled quadrature points in each grid cell where every quadrature has its independent density value. These quadrature points jointly form the scalar field $\mathbf{\rho}$.
In this way, a sub-cell resolution is naturally resolved through multiple quadratures per cell. 
The relation between carrier particles ($\mathbf{x}^c$, $\mathbf{\rho}^c$) and quadrature points ($\mathbf{\rho}$) is further constructed using smoothed-particle hydrodynamics (SPH) kernel and a sharp density mapping function. 
Adopting the SPH kernel has the equivalent effect as the gradient filter \cite{sigmund200199} that prevents the checkerboard pattern. However, since the SPH kernel is directly incorporated in the objective, it avoids performing any extra smoothing on the gradient, keeping the search direction and the objective consistent. 

To efficiently solve the static equilibrium, it is straightforward to achieve a narrow-band sparse simulation \cite{liu2018narrow} using MPM by filtering out low-density quadratures. This is essentially equivalent to how zero-mass grid nodes are filtered out in MPM-based dynamic simulations. 
This mechanism is also adopted to eliminate QR-patterns by maintaining a single main connected component to filter out isolated material blocks. 
The narrow-band filter threshold's increment replaces the Heaviside projection in traditional density-based topology optimization algorithm for producing binarized designs.
\par
\begin{figure}[t]
    \centering
    \includegraphics[width=0.9\textwidth]{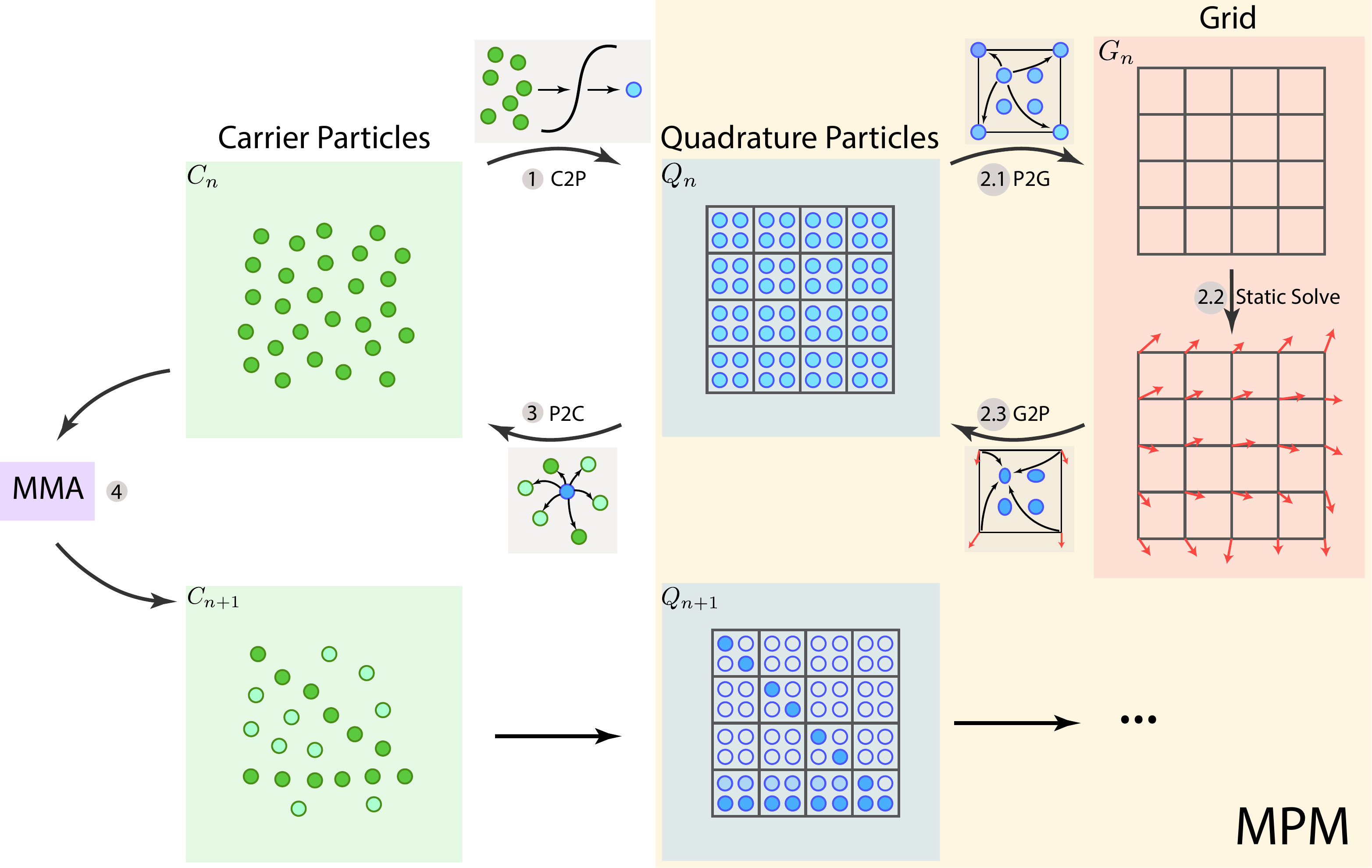}
    \caption{\textbf{Hybrid Lagrangian-Eulerian method pipeline with an MPM solver.}}
    \label{fig:illustration}
\end{figure}

Using moving
asymptotes (MMA) \cite{svanberg1987method} as the optimizer, the optimization pipeline can be summarized as the follows; also see an illustration in \cref{fig:illustration}.

\renewcommand{\labelenumii}{\theenumii}
\renewcommand{\theenumii}{\theenumi.\arabic{enumii}.}
\begin{enumerate}
    \setcounter{enumi}{-1}
    \item \textbf{Initialize:} Collocate carrier particles on the uniformly sampled MPM quadrature points and initialize carrier particle density $\mathbf{\rho}^c$ with the target volume prescription such that the volume constraint is satisfied; see \cref{sec:C2P_transfer}.
    \item \textbf{Transfer information from carrier particles to quadrature points (C2P):} Transfer density from carrier particles to quadrature points ($\mathbf{\rho}(\mathbf{x}^c, \mathbf{\rho}^c)$) with a spherical kernel and a sharp density mapping function; see \cref{sec:C2P_transfer}.
    \item \textbf{MPM Static Solve}; see \cref{sec:mpm_discretize}.
    \begin{enumerate}
        \item \textbf{Transfer information from quadrature points to grid (P2G):} Extract the main connected component using the narraw-band filter; see \cref{sec:narrow-band}. 
        Transfer density from quadrature points in the main connected component to grid nodes and construct the MPM system matrix $\frac{\partial^2 e}{\partial \mathbf{u}^2}$ on the grid. 
        Untouched grids are dropped out of degrees of freedom.
        \item \textbf{Solve force equilibrium:} Solve $\frac{\partial e}{\partial \mathbf{u}} = \mathbf{f}$ for the displacement field $\mathbf{u}$ on the MPM grid subject to Dirichlet boundary conditions $\mathbf{Du} = \mathbf{0}$. Here, only solving a single linear system is required for material with a linear elastic material (see \cref{sec:linear_special_case}). On the other hand, the projected Newton method with line search that guarantees stability and convergence is applied for nonlinearly elastic materials (see \cref{sec:PN}).
        \item \textbf{Update quadrature deformation gradient (G2P):} Update the deformation gradient $\mathbf{F}_q$ of quadrature points with the solved nodal displacement field $\mathbf{u}$.
    \end{enumerate}
    \item \textbf{Compute compliance and the derivatives (P2C):} Evaluate compliance objective $e(\mathbf{\rho}, \mathbf{u})$ (\cref{elastic_potential_compliance}), compliance derivative $de/d\{\mathbf{x}^c,\mathbf{\rho}^c\}$ (\cref{obj_grad}), volume $V$ (\cref{g}), and volume derivative $dV/d\{\mathbf{x}^c,\mathbf{\rho}^c\}$ (\cref{dg_dx}) for optimization search; see \cref{sec:derivative_comp}.
    \item \textbf{Update carrier particle data:} Update $\mathbf{x}^c, \mathbf{\rho}^c$ using MMA and evaluate convergence criteria; see \cref{sec:MMA}. If not converged, go to Step 1 and repeat.
    \item \textbf{Graph-based connectivity correction:} Correct quadrature connectivity according to grid connectivity where the static equilibrium is evaluated; see \cref{sec:narrow-band}.
\end{enumerate}

\section{Hybrid Lagrangian-Eulerian Multi-Density Method}

\subsection{Material Distribution Representation}\label{sec:C2P_transfer}

\begin{figure}[ht]
    \centering
    \begin{subfigure}{0.5\textwidth}
        \includegraphics[width=\textwidth]{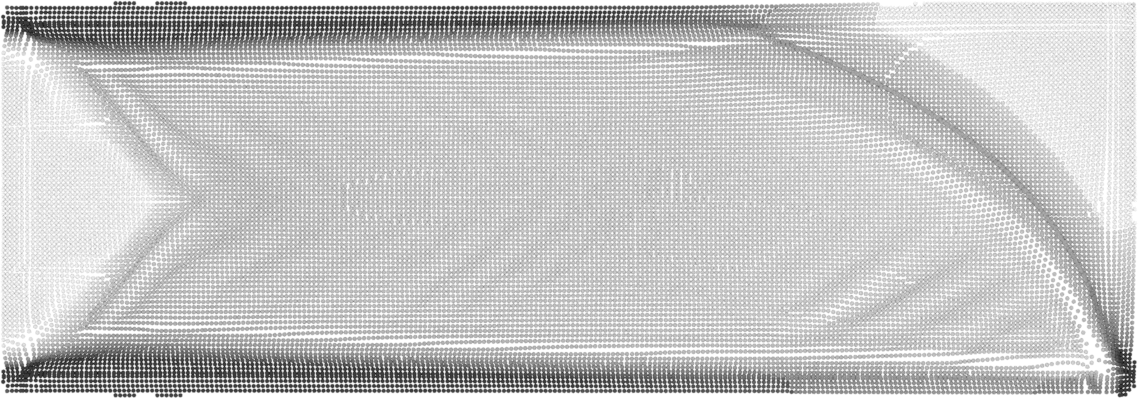}
    \end{subfigure}%
    \begin{subfigure}{0.5\textwidth}
        \includegraphics[width=\textwidth]{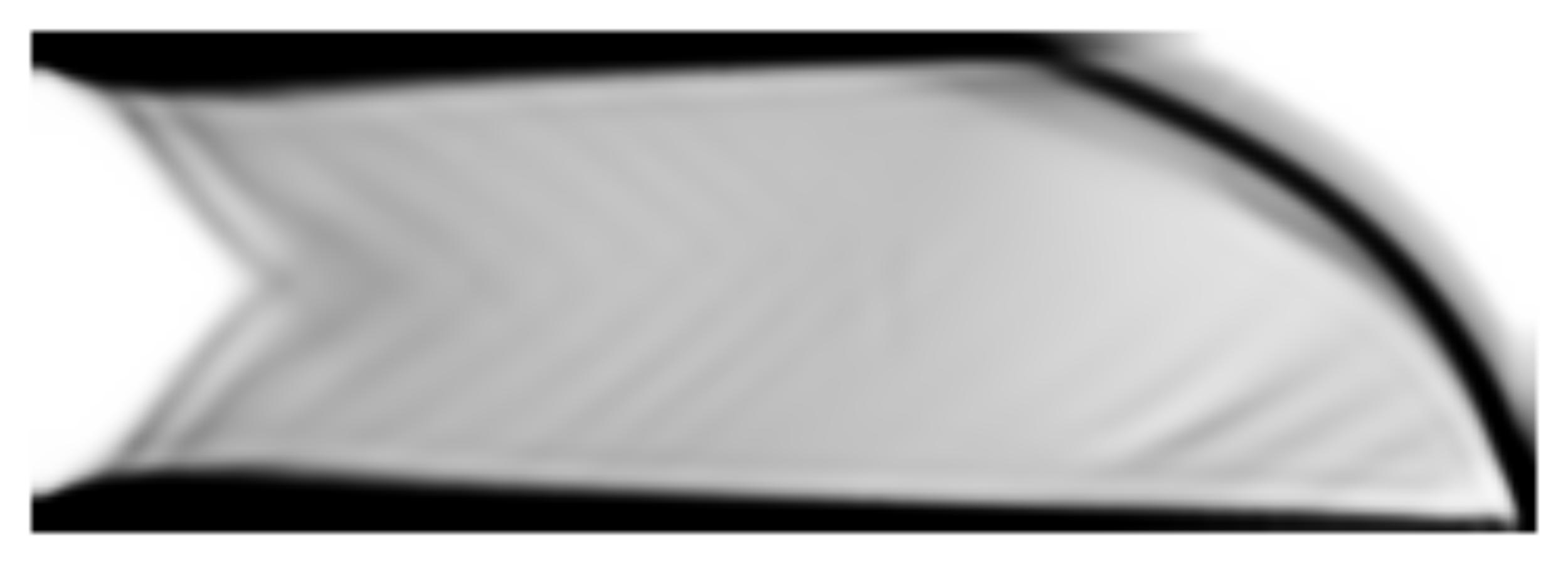}
    \end{subfigure}%
    \\
    \begin{subfigure}{0.5\textwidth}
        \includegraphics[width=\textwidth]{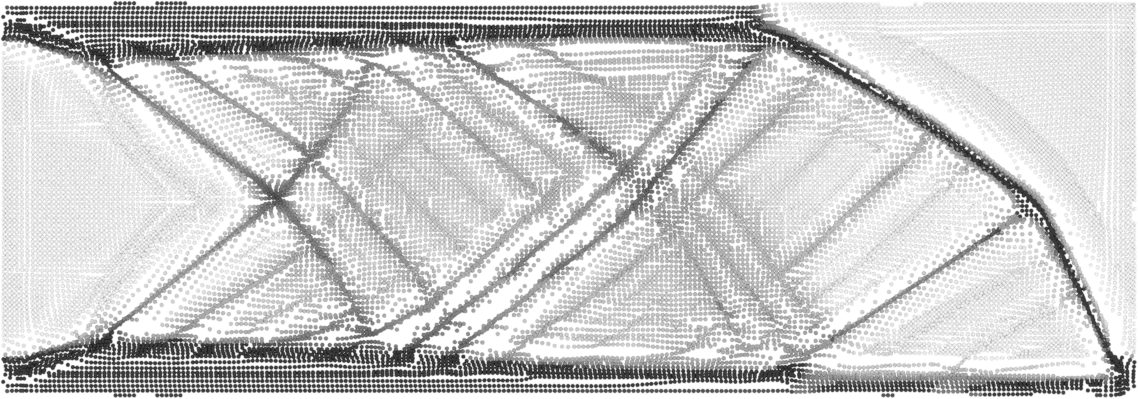}
    \end{subfigure}%
    \begin{subfigure}{0.5\textwidth}
        \includegraphics[width=\textwidth]{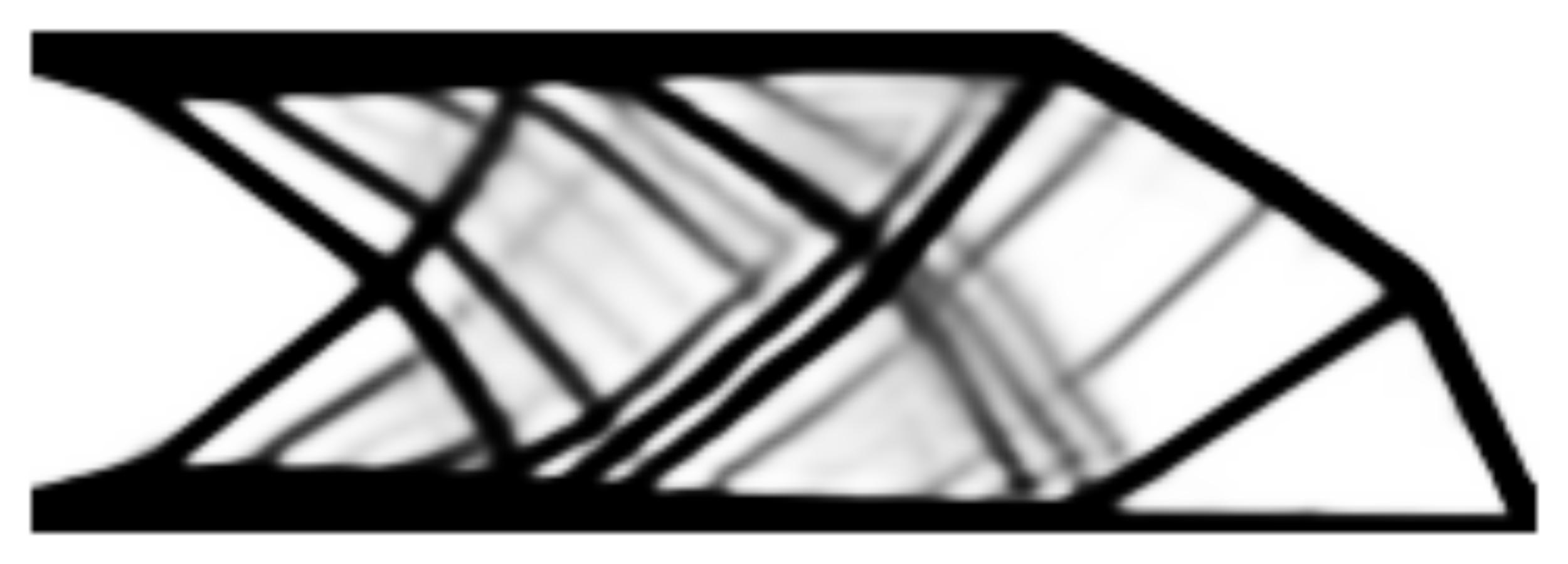}
    \end{subfigure}%
    \\
    \begin{subfigure}{0.5\textwidth}
        \includegraphics[width=\textwidth]{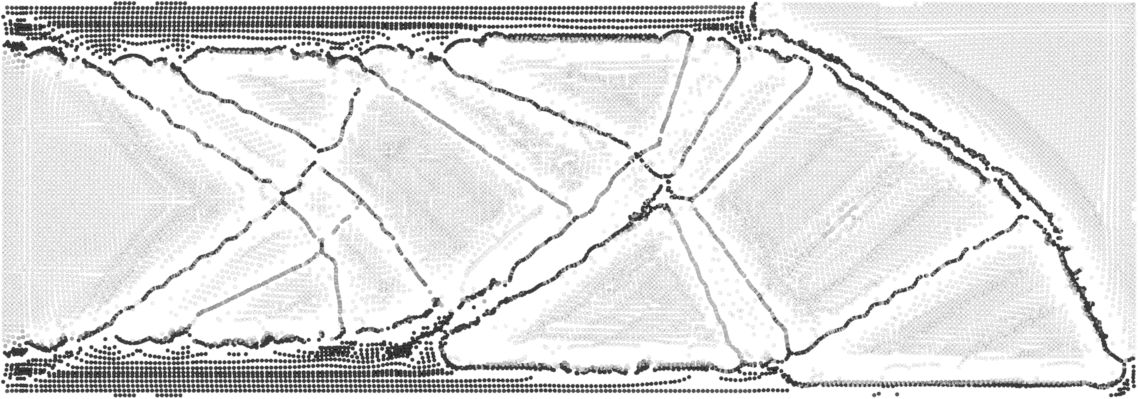}
        \caption{}
    \end{subfigure}%
    \begin{subfigure}{0.5\textwidth}
        \includegraphics[width=\textwidth]{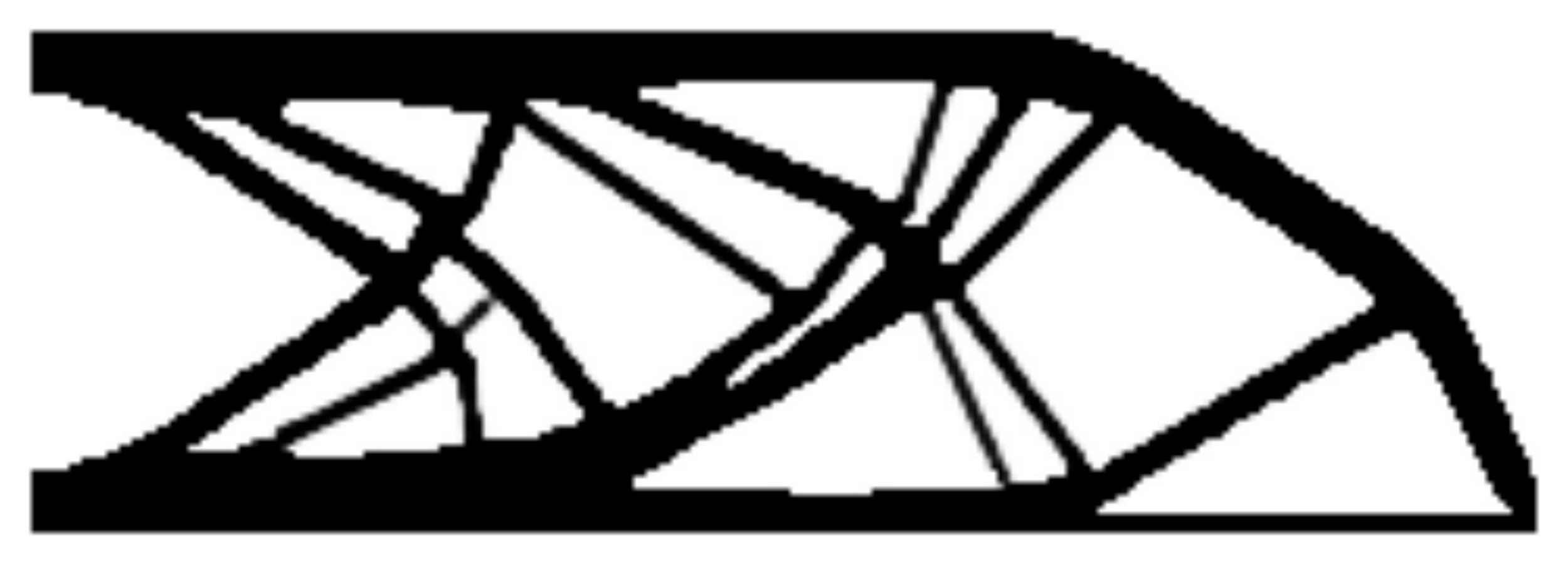}
        \caption{}
    \end{subfigure}%
    \caption{\textbf{Optimization evolution on carrier and quadrature points.} (a) The position and density changes of carrier particles. (b) The density changes on quadrature points.}
    \label{fig:cp_move}
\end{figure}

Introducing Lagrangian degrees of freedom by optimizing quadrature positions together with quadrature densities is a straightforward choice. 
However, arbitrary movements of quadrature may cause large numerical errors, which can even lead to degeneracies like quadrature clustering and isolation.
Thus, another set of moving carrier particles is introduced as the design variables to re-parameterize the density field space and, at the same time, to avoid moving quadrature points.
Carrier particles are defined in the entire design domain with Lagrangian variables $\mathbf{\xi} = (\mathbf{x}^c, \mathbf{\rho}^c)$ consists of both position and density. 
The final material distribution and the volume constraint are still discretized on quadrature points, where their densities are computed according to the surrounding carrier particles. 
In the proposed method, the carrier particles are crucial in emerging intricate geometry structures (see \cref{fig:cp_move}).

The density of each quadrature point $q$ is defined as the weighted sum of its neighboring carrier particles $\{\alpha\}$ using an SPH kernel:
\begin{equation}
    \tilde{\rho}_q = \sum_\alpha \int_\Omega \rho_\alpha^c W\left(\frac{|\mathbf{x}^c_\alpha - \mathbf{X}|}{h}\right) d\mathbf{X} \approx \sum_\alpha \rho_\alpha^c W\left(\frac{|\mathbf{x}^c_\alpha - \mathbf{x}_q|}{h}\right) V_\alpha,
\label{SPHKernel}
\end{equation}
where $W(R)$ is a kernel function, $h$ is the kernel size, and $V_\alpha$ is the volume of each quadrature, which equals to $(\frac{\Delta x}{2})^d$ when $2^d$ quadratures are sampled in each cell. In this paper, the kernel function with cubic spline is applied:
\begin{equation}
    W(R) = \sigma \begin{cases}
    1 - \frac{3}{2} R^2 + \frac{3}{4} R^3, & 0 < R < 1 \\
    \frac{1}{4}(2 - R)^3, & 1 < R < 2 \\
    0, & \text{otherwise}
    \end{cases}
\end{equation}
where $\sigma$ is a constant of $\frac{10}{7\pi h^2}$ in 2D and $\frac{1}{\pi h^3}$ in 3D.

The derivative of $\tilde\rho_q$ w.r.t. the design variables is given by 
\begin{equation}
    \frac{\partial \tilde{\rho}_q}{\partial \mathbf{x}^c_\alpha} = 
        \rho^c_\alpha \frac{\partial W}{\partial \mathbf{x}^c_\alpha} V_\alpha,
\end{equation}
and
\begin{equation}
    \frac{\partial \tilde{\rho}_q}{\partial {\rho}^c_\alpha} = 
        W\left(\frac{|\mathbf{x}^c_\alpha - \mathbf{x}_q|}{h}\right) V_\alpha.
\end{equation}

\begin{figure}[t!]
    \centering
    \includegraphics[width=0.4\textwidth]{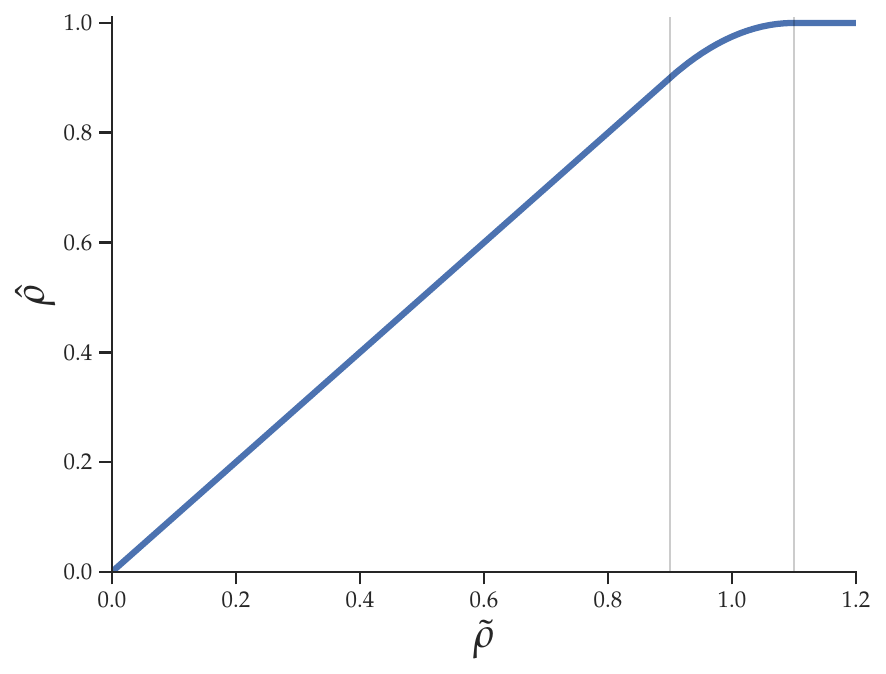}
    \caption{Density mapping function.}
    \label{fig:wrapper}
\end{figure}

To prevent the densities of quadrature particles from exceeding one,
a smooth clamping function is further added on top of $\tilde{\rho}_q$ (see \cref{fig:wrapper}):
\begin{equation}
    \hat{\rho}(\tilde{\rho}) = \begin{cases}
    \tilde{\rho}, & 0 \leq \tilde{\rho} < 1 - \epsilon \\
    \frac{(\tilde{\rho} + \epsilon - 1)^2}{4\epsilon} + \tilde{\rho}, & 1 - \epsilon \leq \tilde{\rho} < 1 + \epsilon\\
    1, & \tilde{\rho} \ge 1 + \epsilon
    \end{cases}
\label{wrapper}
\end{equation}
and its derivative is
\begin{equation}
    \frac{\partial \hat{\rho}}{\partial \tilde{\rho}} = \begin{cases}
    1, & 0 \leq \tilde{\rho} < 1 - \epsilon \\
    \frac{(\tilde{\rho} + \epsilon - 1)}{2\epsilon} + 1, & 1 - \epsilon \leq \tilde{\rho} < 1 + \epsilon\\
    0. & \tilde{\rho} \ge 1 + \epsilon
    \end{cases}.
\label{wrapper_gradient}
\end{equation}

If $\tilde{\rho}$ is greater than $1 + \epsilon$, the output will be one, and the derivative will be zero. Consequently, through chain-rule, the gradients w.r.t. the design variables vanish, which successfully prevent aggregation of particles. Otherwise, stiffer material than allowed can be formed by gathering particles, leading to non-physical material.
\par
Finally, the density of each quadrature $q$ is given by,
\begin{equation}
    \rho_{q} = \hat{\rho}(\tilde{\rho}),
\end{equation}
and its derivative is
\begin{equation}
    \frac{d\rho_q}{d\xi} =  \frac{d\hat{\rho}_q}{d\tilde\rho_q} \frac{d\tilde\rho_q}{d\mathbf{\xi}}.
\end{equation}

The volume of the structure is given by
\begin{equation}
    V({\mathbf{\xi}}) = \sum_{q} \rho_{q},
    \label{g}
\end{equation}
where $q$ indices all quadrature points, and its derivative is
\begin{equation}
    \frac{dV}{d{\mathbf{\xi}}} = \sum_{q} \frac{\partial \rho_{q}}{\partial {\tilde\rho}} \frac{\partial \tilde\rho}{\partial {\mathbf{\xi}}},
    \label{dg_dx}
\end{equation}

\par
The densities of carrier particles are initialized to a uniform scale such that the density of each quadrature is equal to the prescribed volume fraction (the prescribed volume divided by the domain volume).
Since the clamping function is only nonlinear after $1-\epsilon$, by the partition-of-unity property of SPH kernel, the initial volume is very close to the target volume, where the error comes from the approximation in \cref{SPHKernel}. Also note that since the volume constraints are defined on quadrature volumes and the output structure is also represented by quadratures with nonzero density values, there is no need to consider the mass/volume conservation during carrier-quadrature transfers.

\subsection{Design Sensitivity Analysis}\label{sec:derivative_comp}

To compute the derivatives of the compliance objective $e$ w.r.t. design variables ${\mathbf{\xi}}$ required for the topology optimization, the searching process of finding the adjoint variables is constrained to be solely on the force equilibrium manifold.
\par
$\frac{de}{d {\mathbf{\xi}}}$ can be expressed via applying the chain rule as
\begin{equation}
    \frac{de}{d {\mathbf{\xi}}} = \left[\frac{d\mathbf{\rho}}{d\mathbf{\mathbf{\xi}}} \right]^T\left(\frac{\partial e}{\partial \mathbf{\rho}} + \left[\frac{d \mathbf{u}}{d \mathbf{\rho}}\right]^T \frac{\partial e}{\partial \mathbf{u}}\right).
    \label{de_dx}
\end{equation}
Here $\frac{d \mathbf{u}}{d \mathbf{\rho}}$ is difficult to compute as $\mathbf{u}$ and $\mathbf{\rho}$ are related by the force equilibrium equation; even the evaluation of $\mathbf{u}$ from $\mathbf{\rho}$ requires solving a system of equations.
However, if the searching process is constrained to be only on the constraint manifold defined by the force equilibrium equation, differentiating $\frac{\partial e}{\partial \mathbf{u}} = \mathbf{f} $ \wrt $\mathbf{\rho}$ provides
\begin{equation}
    \frac{\partial^2 e}{\partial \mathbf{\rho} \partial \mathbf{u}} + \left[\frac{d \mathbf{u}}{d \mathbf{\rho}}\right]^T \frac{\partial^2 e}{\partial \mathbf{u}^2}  = 0,
    \label{steady_state_diff}
\end{equation}
then \cref{de_dx} can be simplified by combining with \cref{steady_state_diff} into
\begin{equation}
    \frac{de}{d\mathbf{\xi}} = \left[\frac{d\mathbf{\rho}}{d\mathbf{\xi}}\right]^T\left(\frac{\partial e}{\partial \mathbf{\rho}} - \frac{\partial^2 e}{\partial \mathbf{\rho} \partial \mathbf{u}} \left[\frac{\partial^2 e}{\partial \mathbf{u}^2}\right]^{-1} \mathbf{f}\right),
    \label{obj_grad}
\end{equation}
which is the final derivative, where the compliance $e$ can be defined by either linear or nonlinear elasticity.

\paragraph{Simplification Under Linear Elasticity}\label{sec:linear_special_case}
The computation of derivative $\frac{de}{d\mathbf{\rho}}$ requires solving a linear system. However, when linear elasticity is applied, it can be simplified to the form widely used in linear topology optimization \cite{sigmund200199}.
\par
Specifically, when linear elasticity is utilized, the matrix $\frac{\partial^2 e}{\partial \mathbf{u}^2}$ is constant, so the internal elasticity force is linear \wrt $\mathbf{u}$, and the potential $e$ is quadratic \wrt $\mathbf{u}$. Namely, 
\begin{equation}
\begin{split}
    & e = \frac{1}{2} \mathbf{u}^T \mathbf{K} \mathbf{u}, \quad
    \mathbf{f} = \frac{\partial e}{\partial \mathbf{u}} = \mathbf{K} \mathbf{u}, \quad
    \frac{\partial^2 e}{\partial \mathbf{u} \partial \mathbf{u}} = \mathbf{K}, \\
    & \frac{\partial e}{\partial \mathbf{\rho}} = \frac{1}{2} \mathbf{u}^T \frac{\partial \mathbf{K}}{\partial \mathbf{\rho}} \mathbf{u}, \quad
    \frac{\partial^2 e}{\partial \mathbf{\rho} \partial \mathbf{u}} = \mathbf{u}^T\frac{\partial \mathbf{K}}{\partial \mathbf{\rho}},
\end{split}
\end{equation}
where $\mathbf{K}$ is the stiffness matrix depending on densities $\mathbf{\rho}$.
Substituting these equations into \cref{obj_grad}, it follows that
\begin{equation}
    \frac{d e}{d\mathbf{\rho}} = \frac{1}{2} \mathbf{u}^T \frac{\partial \mathbf{K}}{\partial \mathbf{\rho}} \mathbf{u} - \mathbf{u}^T \frac{\partial \mathbf{K}}{\partial \mathbf{\rho}} \mathbf{K}^{-1}\mathbf{f} = -\frac{1}{2} \mathbf{u}^T\frac{\partial \mathbf{K}}{\partial \mathbf{\rho}}\mathbf{u},
\end{equation}
which is exactly the same form derived in traditional linear topology optimization via the adjoint method. For arbitrary design variables $\mathbf{\xi}$, applying the chain-rule, the derivative becomes
\begin{equation}
    \frac{de}{d\mathbf{\xi}}= -\frac{1}{2} \left[\frac{d\mathbf{\rho}}{d\mathbf{\xi}}\right]^T \left[ \mathbf{u}^T\frac{\partial \mathbf{K}}{\partial \mathbf{\rho}}\mathbf{u}\right].
\end{equation}

\subsection{MPM Discretization for Multi-Density Topology Optimization}\label{sec:mpm_discretize}

In MPM, the design domain $\Omega$ is discretized with a set of material particles, or quadratures to approximate the integrals in \cref{eq:formulation}, which are computed as the weighted sum over all the quadratures:
\begin{equation}
    e(\mathbf{\rho}, \mathbf{u}) = \int_{\Omega} \Psi(\mathbf{F}) dX \approx \sum_q  \Psi(\mathbf{F}_q)  V_q.
\end{equation}
\begin{equation}
    V(\mathbf{\rho}) = \int_{\Omega} \rho(\mathbf{X}) dX \approx \sum_q  \rho_q  V_q.
\end{equation}
where $q$ indices all quadrature points, and each quadrature $q$ has its own density $\rho_q$, Young's modulus $E_q$, deformation gradient $F_q$, elastic energy density function $\Psi_q$, and volume $V_q$. To get a sufficiently aligned density field, $2^d$ quadrature points in each cell are sampled on a regular lattice structure as illustrated in \cref{fig:illustration}. With MPM discretization, multiple densities per cell can be handled in a straightforward manner.
\par
Similarly to SIMP, Young's modulus is scaled by the powered density of the particle: $E_q = \rho_q^p E_0$, where $E_0$ is the material's Young's modulus, so that the stiffness of the material is continuously varying over the domain according to its distribution.
Since $\Psi$ is linear \wrt Young's modulus, the compliance can be rewritten as
\begin{equation}
    e(\mathbf{\rho}, \mathbf{u}) = \sum_q \rho_q^p \Psi_0(\mathbf{F}_q) V_q,
    \label{elastic_potential_compliance}
\end{equation}
where $\Psi_0$ is the energy density function with Young's modulus $E_0$.

\paragraph{Static Equilibrium}
While MPM is commonly used for dynamic time-stepping, a static
problem formulation based only on MPM spatial discretization similar to the total Lagrangian formulation\ \cite{de2020total} is needed for topology optimization. In such a static setting, there is no inertia effect or time-variant variables, meaning that only the elasticity and external force terms are kept:
\begin{equation}
    -\frac{\partial \mathbf{e}}{\partial \mathbf{u}}(\mathbf{\rho}, \mathbf{u}) + \mathbf{f} = \mathbf{0},
    \label{eq:staticMPM}
\end{equation}
where $\mathbf{f}$ is the external body force load (Neumann boundary condition) defined on the grid nodes. Dirichlet boundary conditions can be defined as
\begin{equation}
    \mathbf{Du} = \mathbf{0},
    \label{eq:DBC}
\end{equation}
where $\mathbf{D}$ is the selection matrix that extracts the Dirichlet grid nodes. From a variational point of view, this is equivalent to solving the following optimization problem
\begin{equation}
    \min_{\mathbf{u}} e(\mathbf{\rho}, \mathbf{u}) - \mathbf{u}^T \mathbf{f} \quad \text{s.t.} \quad \mathbf{Du} = \mathbf{0}.
    \label{eq:staticMin}
\end{equation}
\par
In MPM, quadratures are embedded in the background Eulerian grid with a B-spline kernel, meaning that the nodal displacement $\mathbf{u}$ is, in fact, defined on the uniform grid nodes. Since in \LETO elastostatic problems are solved without material particle (quadrature points) advection, there will be no cell crossing errors or ringing instabilities for MPM even if linear kernel is used for simplicity:
\begin{equation}
    N(x) = \begin{cases}
    1 - |x|, & 0 \leq x < 1 \\
    0, & 1 \leq x.
    \end{cases}
\end{equation}
Here the weight $\omega_{iq}$ between grid node location $\textbf{x}_i$  and quadrature location $\textbf{x}_q$ is defined by taking the Cartersian product in all dimensions. For example, in 3D, 
\begin{equation}
    \omega_{iq}(x_q) = N(\frac{1}{h}(x_{q,1} - x_{i,1}))N(\frac{1}{h}(x_{q,2} - x_{i,2}))N(\frac{1}{h}(x_{q,3} - x_{i,3}));
\end{equation}
and in 2D, 
\begin{equation}
    \omega_{iq}(x_q) = N(\frac{1}{h}(x_{q,1} - x_{i,1}))N(\frac{1}{h}(x_{q,2} - x_{i,2})).
\end{equation}
The deformation gradient $F_q$ on quadrature $q$ is then related to the surrounding grid nodes $i$ as
\begin{equation}
    \mathbf{F}_q = \mathbf{I} + \sum_i \mathbf{u}_i \nabla \mathbf{\omega}_{iq}^T,
\end{equation}
which also leads to the elasticity force definition
\begin{equation}
    -\frac{\partial \mathbf{e}}{\partial \mathbf{u}} = -\sum_q \rho_q^p V^0_q \frac{\partial \Psi_0(\mathbf{F}_q)}{\partial \mathbf{F}_q} \nabla \mathbf{\omega}_{iq},
\end{equation}
and the elasticity Hessian (in index notation)
\begin{equation}
    \frac{\partial^2 e}{\partial u_{i,\alpha} \partial u_{j,\beta}} = \sum_q \rho_q^p V^0_q (\nabla \omega_{iq})_\delta \frac{\partial^2 \Psi_0(F_q)}{\partial F_{q,\alpha \delta} \partial F_{q,\beta\omega}} (\nabla \omega_{jq})_\omega,
\end{equation}
where $1 \leq \alpha, \beta, \delta, \omega \leq d$ and $d$ is the spatial dimension. $F_{q,\alpha \beta}$ is $(\alpha, \beta)$-th element of the deformation gradient $F_q$.
\par
Compared with the MPM formulation for dynamic problems, the static formulation can be seen as only solving for a single ``time step,'' and the deformation gradient at previous time step, $\mathbf{F}^n_q$, is just the initial undeformed deformation gradient $ \mathbf{F}^n_q = \mathbf{F}^0_q = \mathbf{I}$.

\paragraph{Static Solve with Projected Newton}\label{sec:PN}
To solve the equilibrium equation more robustly, the variational form (\cref{eq:staticMin}) is minimized with the projected Newton's method \cite{teran2005robust} as outlined in \cref{alg:newton}. 
Dirichlet boundary conditions are handled by modifying the corresponding entries in the matrix and the right-hand-side to keep the problem unconstrained, which is equivalent to eliminating the Lagrange multipliers in the KKT system with linear equality constraints. 
The stopping criteria is chosen to be $||\Delta \mathbf{u}||_{\infty} < \tau = 0.1\Delta \mathbf{x}$. Note that when a linear constitutive model is used, the system is quadratic, so only one iteration is needed.

\algdef{SE}[DOWHILE]{Do}{doWhile}{\algorithmicdo}[1]{\algorithmicwhile\ #1}

\begin{algorithm}
    \caption{Projected Newton for Solving Static Equilibrium}
    \label{alg:newton}
    \begin{algorithmic}[1]
    \Procedure{ProjectedNewton}{$\rho^j$, $\mathbf{f}$, $\mathbf{D}$, $\tau$, $\mathbf{u}$}
        \State $\Delta \mathbf{u} = \mathbf{0}, \mathbf{u}^0 = \mathbf{0}, i = 0$ \hspace{10pt} // initialize
        \Do
            \State $\mathbf{P} \leftarrow \text{projectSPD}(\frac{\partial^2 e}{\partial \mathbf{u}^2})$ \hspace{10pt} // project each local Hessian stencil to SPD\ \cite{teran2005robust}
            \State $\Delta \mathbf{u}\leftarrow$ $\mathbf{P}^{-1}(\mathbf{f} - \frac{\partial e}{\partial \mathbf{u}})$
            \State $\alpha$ $\leftarrow$ LineSearch($\mathbf{u}^i, \Delta \mathbf{u}$) \hspace{10pt} // Back-tracking line-search
            \State $\mathbf{u}^{i+1} \leftarrow$ $\mathbf{u}^i + \alpha \Delta \mathbf{u}$
            \State $i \leftarrow i + 1$
        \doWhile{$||\Delta \mathbf{u}||_{\infty} \geq \tau$}
        \State $\mathbf{u} \leftarrow \mathbf{u}^i$
    \EndProcedure
    \end{algorithmic}
\end{algorithm}

\paragraph{Inversion-free Line Search}
Since in each projected Newton iteration, the Hessian has been projected to symmetric positive definite, the search direction $\Delta \mathbf{u}$ is guaranteed to be a descent direction. Therefore, back-tracking line search can ensure $E(\mathbf{u}^{i+1}) < E(\mathbf{u}^i)$ after each $\mathbf{u}$ update, which effectively stabilizes the iterations and improves convergence.

However, for the noninvertible elasticity energy (neo-Hookean), projected Newton does not necessarily ensure no deformation gradient inversion along search direction $\Delta \mathbf{u}$. Hence, following Smith and Schaefer \cite{smith2015bijective}, to further prevent inversion of each $\mathbf{F}_q$, a large feasible step size before each line search is solved by finding the minimum of the smallest positive roots of a family of equations
\begin{equation}
    \left\{\det({\mathbf{F}_q(\mathbf{u}^i + \beta_q \Delta \mathbf{u})}) = \epsilon_q\right\}.
\end{equation}
The line search step size then starts from $\min_q \beta_q$. Here, $\epsilon_q=0.1\det(\mathbf{F}_q(\mathbf{u}^i))$ is used to avoid numerical rounding errors, which is more robust than solving with $\epsilon_q=0$.

\subsection{Narrow-band Filter and Connectivity Correction} \label{sec:narrow-band}
Multi-density topology optimization methods suffer from a common artifact known as QR-patterns. 
Unlike the traditional checkerboard problem, QR-patterns happen on a sub-cell level, which corresponds to the quadrature here. 
When QR-patterns appear, there are many sub-cell level isolation of the solid components, which are still viewed as connected from the perspective of the background grid where forces and displacements are discretized.
This kind of inconsistency prevents simulation from accurately predicting a structure's compliance, which can lead to results with large compliance when tested in practice or simulated on a grid with higher resolution.
Based on this observation, this paper proposes a narrow-band filter to keep only one major connected component during the optimization and then apply a connectivity correction step to correct the final topology further.
\par
\begin{figure}[ht]
    \centering
    \includegraphics[width=0.4\textwidth]{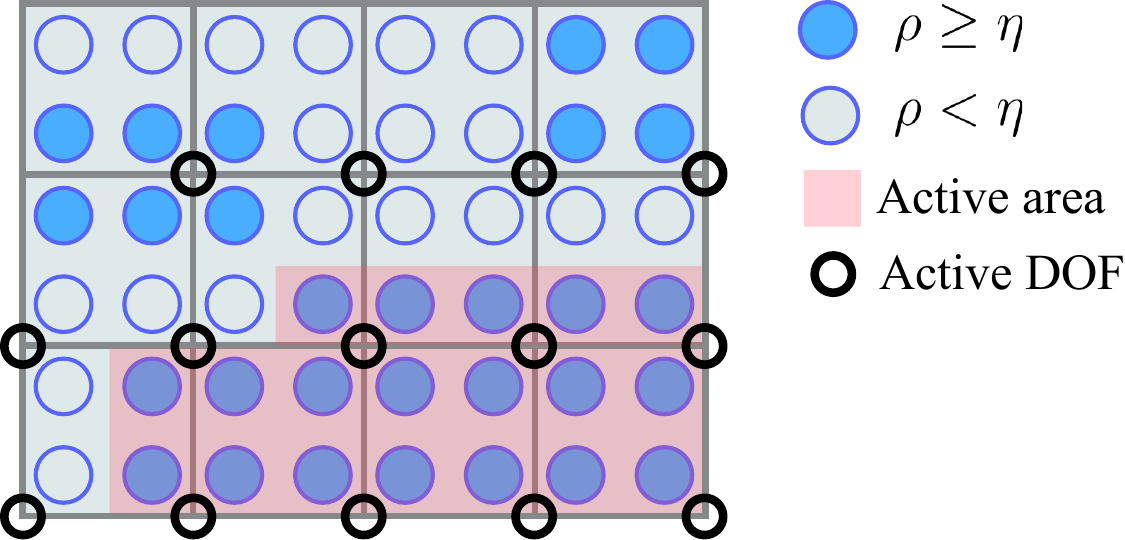}
    \caption{An illustration of the narrowband filter mechanism. An active area is first extracted by the graph connectivity. Active DOFs are then indicated by whether they belongs to cells with active quadratures. Only active DOFs are kept for the static solve to reduce the computational cost.} 
    \label{fig:narrowband-fitler-illustration}
\end{figure}

Given a threshold $\eta$, a graph is built at each optimization iteration before the static solve, with its vertices being those quadrature points having a density greater than $\eta$. 
Quadrature pairs are identified to be adjacent only if their Manhattan distance is $1$ (adjacent along one of the coordinate axes). 
By performing a breadth-first search on this graph, the connected component $\Theta$ with the most quadrature points are then extracted. Only the grid nodes within the kernel range of active quadrature points are kept as DOFs in the MPM static solve step. An illustration is shown in \cref{fig:narrowband-fitler-illustration}.

The narrowband filter $\theta$ employed is introduced as:
\begin{equation}
    \theta(\rho_q) = 
    \begin{cases}
     0, \ \ &q \in \Theta \\
     1, \ \ &q \not\in \Theta,
    \end{cases}
\end{equation}
with which the topology optimization becomes:
\begin{equation}
    \underset{\mathbf{\xi}, \mathbf{u}}{\text{min}}\:\: e(\mathbf{\rho}, \mathbf{u}) = \sum_q \theta(\rho_q) \rho_q^p \Psi_0(\mathbf{F}_q) V_q,
    \quad \text{s.t.} \quad
    \begin{cases}
        \frac{\partial e}{\partial \mathbf{u}}(\mathbf{\rho}, \mathbf{u}) = \mathbf{f} \\
        \mathbf{Du} = \mathbf{0}\\
        V(\mathbf{\rho}) \le \hat{V}.
    \end{cases}
\end{equation}
To handle the non-smoothness of $\theta$,an alternating optimization style strategy is performed that computes $\theta$ between each MMA optimization iteration where $\theta$ is then treated as constant. Theoretically, this strategy can lead to different local optimum compared to a fully coupled search, but it has been shown to be effective in our experiments.
During the optimization, the narrow-band filter can also accelerate the pruning of redundant branches, for example, a fiber with only one end connected to the major component, which can increase the compliance.
\par
\begin{figure}
    \centering
    \includegraphics[width=0.4\textwidth]{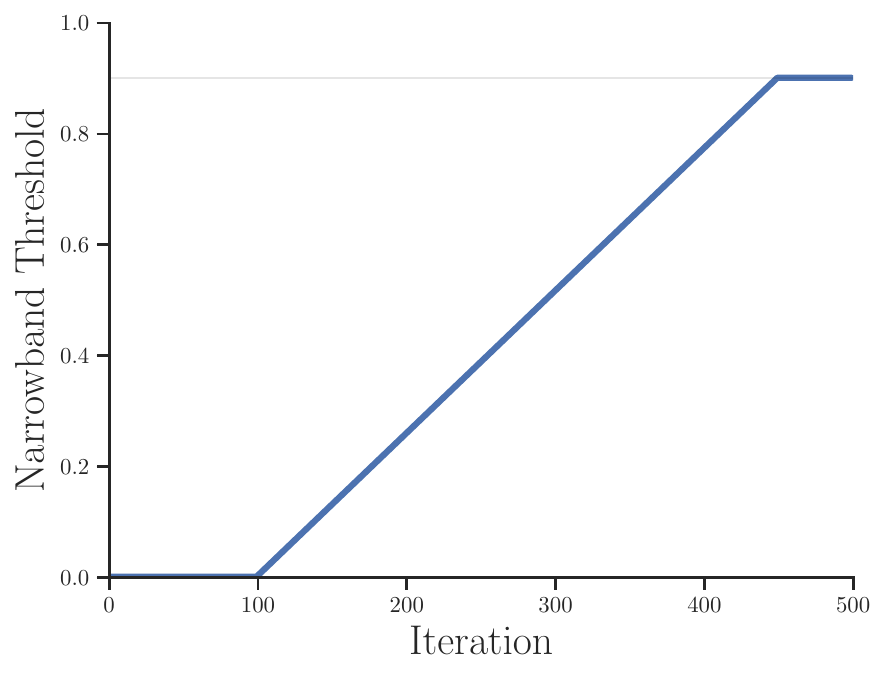}
    \caption{Evolution of narrowband threshold.}
    \label{fig:narrowband_threshold}
\end{figure}

Given that the narrow-band filter can remove all quadrature points with a density below $\eta$, a new binary design enforcement mechanism is proposed. 
A low threshold of the narrow-band filter is chosen initially for a coarse but more global result and is further increased per iteration towards a value very close to $1$ (\cref{fig:narrowband_threshold}).
This mechanism provides better optimization stability by enforcing binary design without any nonlinear projections, e.g., a Heaviside function.
\par
\begin{figure}
    \centering
    \begin{subfigure}{0.4\textwidth}
        \includegraphics[width=\textwidth]{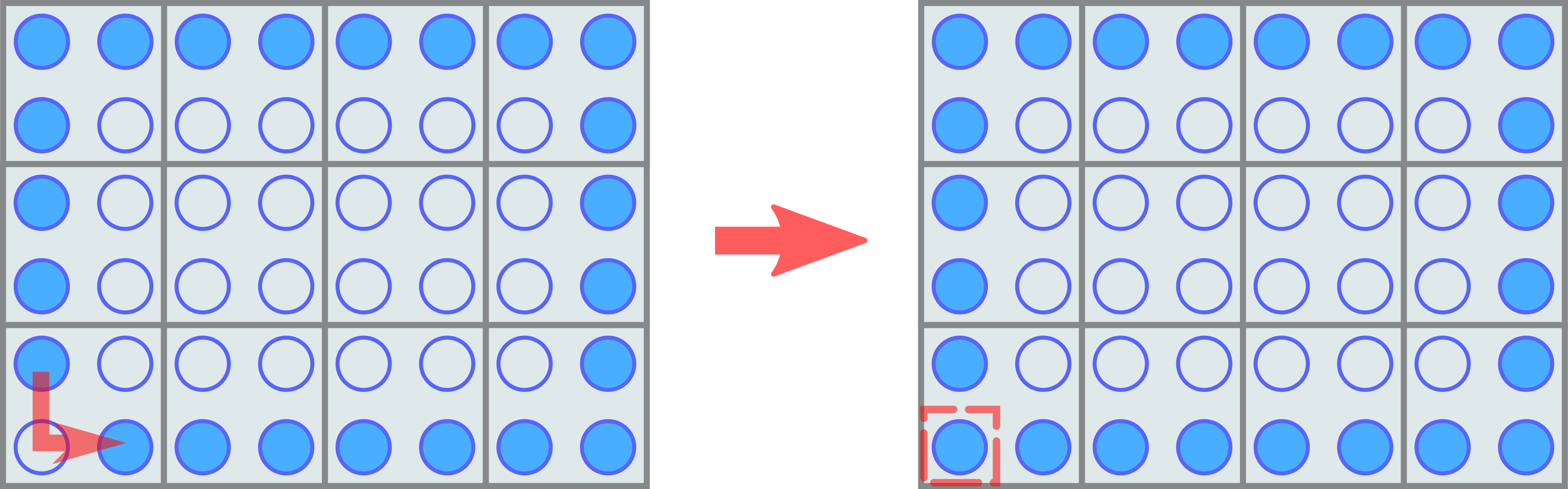}
    \end{subfigure}\ \ \ \ \ \ \ \ \ \ \ \ \ \  
    \begin{subfigure}{0.4\textwidth}
        \includegraphics[width=\textwidth]{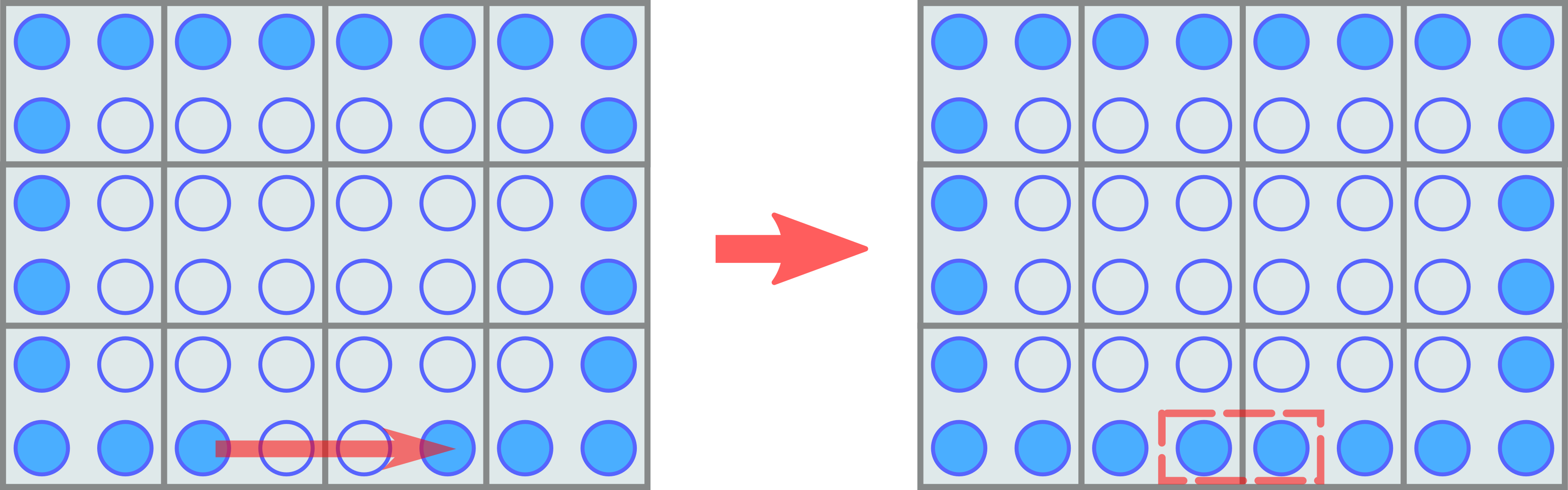}
    \end{subfigure}
    \caption{Illustrated examples of emerging QR-patterns and connectivity correction procedures. }
    \label{fig:qr-pattern}
\end{figure}

However, QR-patterns may still exist when the two parts of the major component's boundaries are spatially close.
\cref{fig:qr-pattern} lists some illustrated examples. Here the filled and unfilled circles are used to discriminate solid and void materials. 
It can be seen that despite a notable disconnection of quadrature points, when evaluating the static force equilibrium, these are still considered connected since they can belong to the same cell (the left example) or adjacent cells (the right example). 
However, when evaluating the resulting design, a double refined grid where each quadrature corresponds to a different cell, the corresponding cells of these separated quadratures will only share one node or are completely disconnected.
To detect and remove these QR-patterns, a connectivity correction step is further introduced: 
for any pair of quadratures belonging to the same cell or two different cells sharing at least one node, 
if the difference between their distance on the graph (the length of the shortest path) and their spatial $L_1$ distance is larger than a threshold, 
a shortest Manhattan path will be created between them with the densities along the path all set to $1$. 
Note that the occurrence of such corrections is rare.

\section{Numerical Examples}
\subsection{Optimizing Structures with MMA} \label{sec:MMA}
The method of moving
asymptotes (MMA) \cite{svanberg1987method} is used to solve the optimization problem, which jointly optimizes the positions and densities of carrier particles.
This optimizer is designed for general structural optimization problems with inequality constraints and box constraints. 
The algorithm approximates the original problem with a series of separable convex optimizations. At each iteration, it sets up two asymptotes for each variable to constrain the searching interval. These asymptotes will be updated according to each sub-optimum. 
In our implementation, we adopt an open-source C++ version of MMA.\footnote{\url{https://github.com/jdumas/mma}}
\par
To obtain high-quality results from MMA, careful parameter tuning is necessary, and different examples might have different sets of optimal MMA parameters. 
There are three parameters to tune: $asyinit, asyincr$, and $asydecr$. In this paper, these parameters are set to be $0.02, 1.05$, and $0.65$, respectively, as in the original MMC method.
The parameters are used throughout all the numerical examples. To further stabilize and accelerate the optimization in a consistent way, the following regularizations are additionally applied within MMA :
\begin{enumerate}
    \item The step length of each variable is controlled by modifying its box constraint at each iteration. The change of carrier density is controlled to be below 0.5, and the change of carrier position at each dimension is controlled to be below 2 times the background grid spacing. This step size control can stabilize optimization.
    \item Following the MMC method, the gradient of objective and the volume constraint are scaled such that their $L^{\infty}$-norms are both 1. In addition, the objective and the volume constraint are scaled accordingly to make sure the scaled gradients are consistent. The scaling accelerates the optimization significantly.
\end{enumerate}

\subsection{Linear Topology Optimization}
\begin{figure}[ht]
\centering
\includegraphics[width=\textwidth]{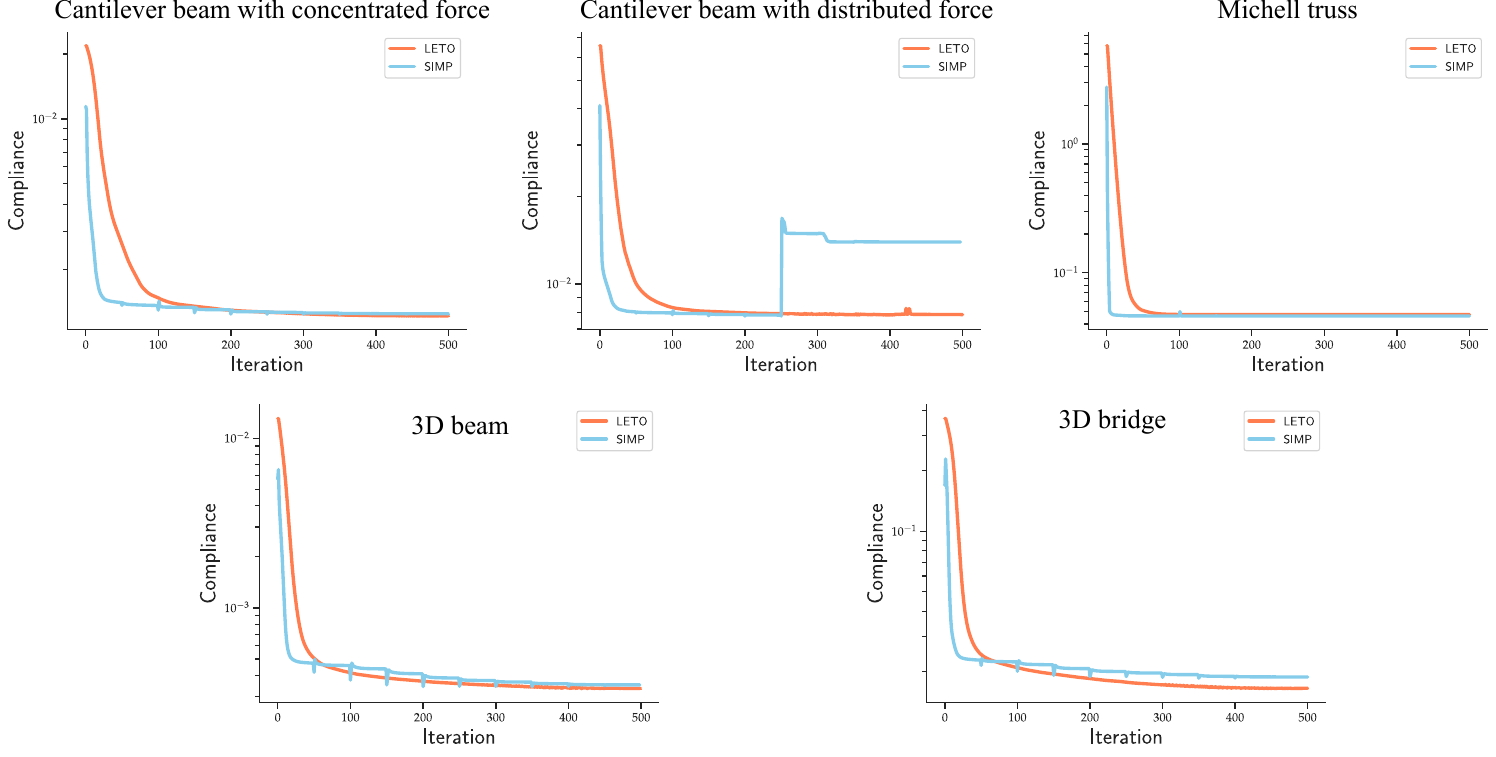}
    \caption{Convergence plots for linear elasticity experiments. 
    It can be seen here in the 5 experiments conducted, \LETO achieves comparable structural compliance value and convergence speed. 
    Notably, under distributed force condition, \LETO delivers better stability.  }
\label{fig:compliance-plot}
\end{figure}

In this section, the proposed method is compared with SIMP with Heaviside projection\cite{guest2004achieving} on linear elasticity examples. The filter radius of SIMP is 1.5 for all examples. 
Under the same simulation resolution, \LETO obtains more detailed geometry structures and delivers comparable or even lower compliance at a similar convergence speed (see \cref{fig:compliance-plot}). 
For fair comparisons, LETO and SIMP's final results are compared on a double refined grid using FEM, where individual quadrature of \LETO corresponds to a cell on the refined grid.
Each cell of SIMP is mapped to a $2\times 2$ (in 2D) or $2 \times 2 \times 2$ (in 3D) cell block with the same density. 
As varying volumes can easily lead to compliance changes, to further ensure fairness, the final volume constraint of \LETO is controlled to be slightly lower than the corresponding volume in the SIMP method.
For each of the following experiments, SIMP is first tested with the target volume. The volume constraint of \LETO is then set to be less than the volume of the binarized SIMP's result. Since \LETO treats volume constraints as inequalities through MMA, it usually ends up with an even slightly smaller volume.
The binarization threshold of SIMP is set to be 0.5. Since \LETO can guarantee there exists no density values between 0 and 0.9, the binarization threshold of \LETO is chosen to be 0.9. 
To demonstrate that QR-patterns do not appear in the proposed method's final results, compliance values are evaluated at both the resolution used in optimization and with double-resolution. 
A collection of resulting compliance value and volume percentage of 5 experiments is shown in \cref{tab:linear_table}. The average computational costs per iteration are also shown. Since MMA used in \LETO is usually slower than OC used in SIMP, it is expected that \LETO is slower than SIMP. However,  the narrowband filter can prune inactive DOFs, which can reduce the computational cost a lot after the structure becomes sufficiently binarized (especially in 3D).

\begin{table}[ht]
\centering
\begin{tabular}{c|cccc|ccc}
\hline
\multirow{2}{*}{\begin{tabular}[c]{@{}c@{}}~~\\Experiment\end{tabular}} & \multicolumn{4}{c|}{LETO} & \multicolumn{3}{c}{SIMP (r=1.5)} \\ \cline{2-8} 
 & Compliance & \begin{tabular}[c]{@{}c@{}}\textbf{Compliance}\\\textbf{(refined grid)}\end{tabular} & Volume &\begin{tabular}[c]{@{}c@{}}{Ave. Cost}\\{(per iter.)}\end{tabular} & \begin{tabular}[c]{@{}c@{}}\textbf{Compliance}\\\textbf{(refined grid)}\end{tabular} & Volume & \begin{tabular}[c]{@{}c@{}}{Ave. Cost}\\{(per iter.)}\end{tabular}  \\ \hline
Concentrated-load beam & $1.218 \times 10^{-3}$ & $1.243 \times 10^{-3}$ & 29.9\%& 0.405s & $1.264 \times 10^{-3}$ & 30.1\% & 0.387s\\
Michell truss & $4.736 \times 10^{-2}$ & $5.054 \times 10^{-2}$ & 20.1\% & 0.544s & $5.055 \times 10^{-2}$ & 20.2\%&  0.355s\\
Distributed-load beam & $7.845 \times 10^{-3}$ & $8.386 \times 10^{-3}$ & 39.7\% & 0.867s & $1.658 \times 10^{-2}$ & 40.0\%&  0.408s\\
3D beam & $3.323 \times 10^{-4}$ & $3.334 \times 10^{-4}$ & 19.6\% & 3.753s  & $3.527 \times 10^{-4}$ & 20.0\% & 6.110s\\
3D bridge & $1.653 \times 10^{-2}$ & $1.643 \times 10^{-2}$ & 19.4\% &6.893s & $1.892 \times 10^{-2}$ & 20.0\% & 16.816s\\ \hline
\end{tabular}
    \caption{
    Compliance value and volume percentage for linear elastic experiments.
    The compliance value of results from the proposed method under optimization grid resolution, under double-refined grid resolution and SIMP method under double-refined grid resolution is shown in this table.
        }
\label{tab:linear_table}
\end{table}

\subsubsection{2D Beam with a Concentrated Load}
\label{sec:concentrated_load_2d}
\begin{figure}[ht]
\centering
\includegraphics[width=0.9\textwidth]{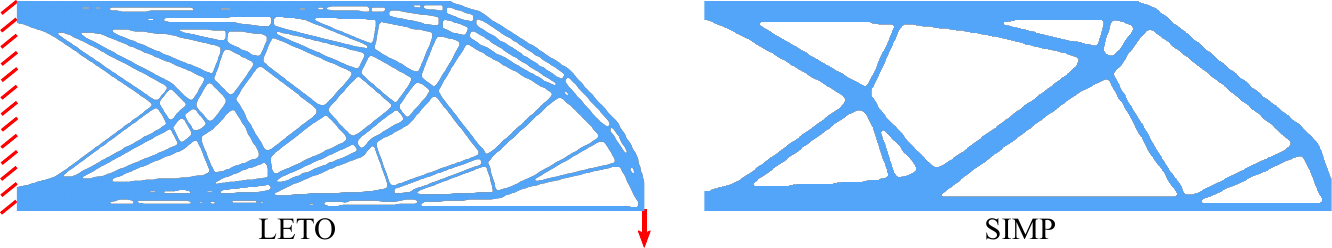}
    \caption{\textbf{2D beam with a concentrated load.}
    The comparison between \LETO and SIMP on this standard 2D beam example where left most grid nodes are fixed, and a concentrated load is applied at the bottom-right corner.
    The compliance value evaluated at double-refined grid resolution for \LETO and SIMP are $1.243 \times 10^{-3}$ and $1.264 \times 10^{-3}$ respectively.
    }
\label{fig:beam}
\end{figure}

In this example, a standard 2D beam benchmark problem is tested; see \cref{fig:beam}. The rectangular design domain is $1m$ in width and $3m$ in length, discretized by a $300 \times 100$ grid with a spacing of $0.01m$. A 0.1N downward force is concentrated at the bottom-right grid node (denoted by the red arrow) and the leftmost grid nodes are fixed. The target volume constraint is 30\%. The optimal material distribution obtained from \LETO has rich branching fibers. Even the thick fibers on the boundary of SIMP's result splits into several thin fibers in \LETO. The compliance of \LETON's result evaluated at optimization resolution and double resolution, and SIMP's result evaluated at double resolution are $1.218 \times 10^{-3}$, $1.243 \times 10^{-3}$, and $1.264 \times 10^{-3}$ respectively. The final volume reached by \LETO and SIMP are 29.9\% and 30.1\%.

\subsubsection{Michell Truss}

\begin{figure}[ht]
\centering
\includegraphics[width=0.75\textwidth]{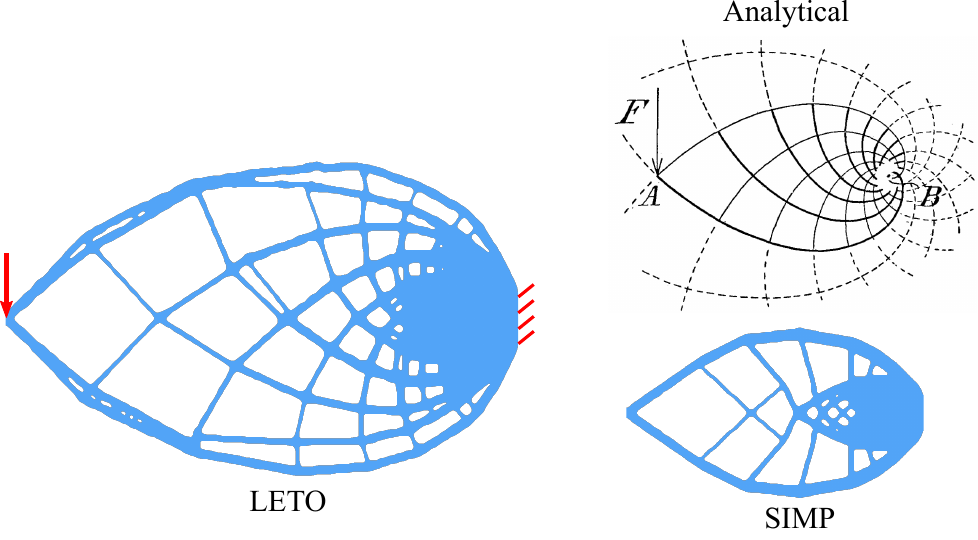}
    \caption{\textbf{Michell truss.}  
    Here the resulting structure of \LETO and SIMP for Michell Truss is evaluated. The central region of the right-most grid is fixed, and a concentrated force is applied in the middle of the leftmost grid.
    The compliance value evaluated at double-refined grid resolution for \LETO and SIMP is $5.054 \times 10^{-2}$ and $5.055 \times 10^{-2}$ respectively.
    The analytical solution \cite{Michell1904} is shown for reference, albeit the compliance value of both methods are not significantly different, the fiber directions from the proposed method align with those appeared in the analytical solution better.}
\label{fig:michell-truss}
\end{figure}

The second example under consideration is Michell truss; see \cref{fig:michell-truss}. A $2m\times1.6m$ rectangle is used as the design domain. The grid resolution is $200 \times 160$ with a spacing of $0.01m$. A concentrated downward force of $1N$ is applied to the middle node of the leftmost column of the grid. The middle region of the right-most grid is set to be fixed. The target volume is 20\%. The compliance of SIMP evaluated at double resolution, \LETO evaluated at optimization resolution, and double resolution are $5.055 \times 10^{-2}$, $4.736 \times 10^{-2}$, $5.054 \times 10^{-2}$ respectively. Although \LETON's result has almost the same compliance as SIMP, it resembles more to the analytical solution \cite{Michell1904} as can be seen from the resulting fiber direction. The final volume reached by SIMP and \LETO is 20.2\% and 20.1\%.

\subsubsection{2D Beam with a Distributed Load}
\begin{figure}[ht]
\centering
\includegraphics[width=\textwidth]{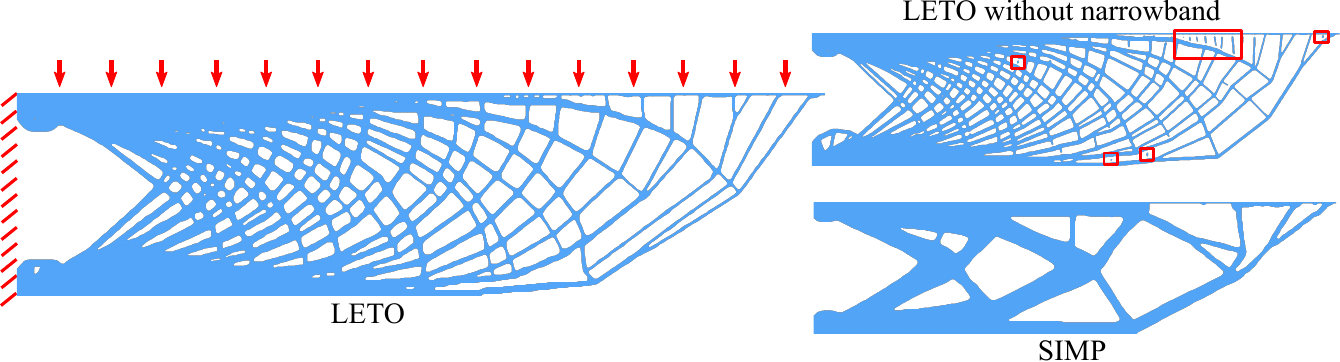}
    \caption{\textbf{2D beam with a distributed load.}
    A distributed force load is applied to this beam example at the top plane where the leftmost nodes are fixed. 
    The compliance value evaluated at double-refined grid resolution for \LETO and SIMP are $8.386\times 10^{-3}$ and $1.658\times 10^{-2}$ respectively.
    \LETO without a narrow-band filter is visualized in the top-right corner to demonstrate that the isolated blobs can be removed with the proposed filtering technique.
    }
\label{fig:distributed-force}
\end{figure}

In this example, the proposed version of \LETO and \LETO without narrow-band filter (simulate with weak material filling across the whole domain) are also compared to demonstrate the capability of the proposed narrow-band mechanism to remove QR patterns; see \cref{fig:distributed-force}. 
To enforce binary design when threshold-increasing narrow-band filter is absent, a family of density mapping functions are used like Heaviside projection in SIMP, except that the functions are smoothly clampped at 0 and 1 to have zero gradient:
\begin{equation}
    \hat{\rho}_k(\tilde{\rho}) = \begin{cases}
    \frac{1}{2}(2 \tilde{\rho})^k, &0 \leq \tilde{\rho} < \frac{1}{2}\\
    1 - \frac{1}{2} (2 - 2\tilde{\rho})^k, &\frac{1}{2} < \tilde{\rho} < 1\\
    1, & \tilde{\rho} \ge 1,
    \end{cases}
\end{equation}
where $k$ is gradually increased from 1.01 to 10 during the optimization.
\par
The design domain is a $4m\times 1m$ rectangle. The left boundary is fixed and a total force of $4N$ is evenly distributed on the top boundary. The grid resolution is $400\times 100$ with a spacing $ 0.01m$. 
The target volume is 40\%. As highlighted by red boxes, a lot of isolated material blobs form when \LETO is used without the narrow-band filter.
Such visually isolated material blobs actually belong to one continuum at the simulation resolution but are disconnected at quadrature level (or higher resolution). 
The large compliance difference between evaluations at low resolution ($c=6.016 \times 10^{-2}$) and high resolution ($c=7.804 \times 10^{-3}$) also indicates the inconsistency, which differ by an order of magnitude. 
The proposed version of \LETO can remove those isolated blobs automatically during the optimization, resulting in consistent compliance values when evaluated at optimization resolution $c=7.845 \times 10^{-3}$ and double resolution ($c=8.386\times 10^{-3}$).
The compliance of SIMP at double resolution is $1.658\times 10^{-2}$. The final volume reached by SIMP and \LETO is 40.0\% and 39.7\%. Results generated by \LETO contains richer intricate fibers than the one by SIMP.

\subsubsection{3D Beam}
\begin{figure}[ht]
    \centering
    \includegraphics[width=0.9\textwidth]{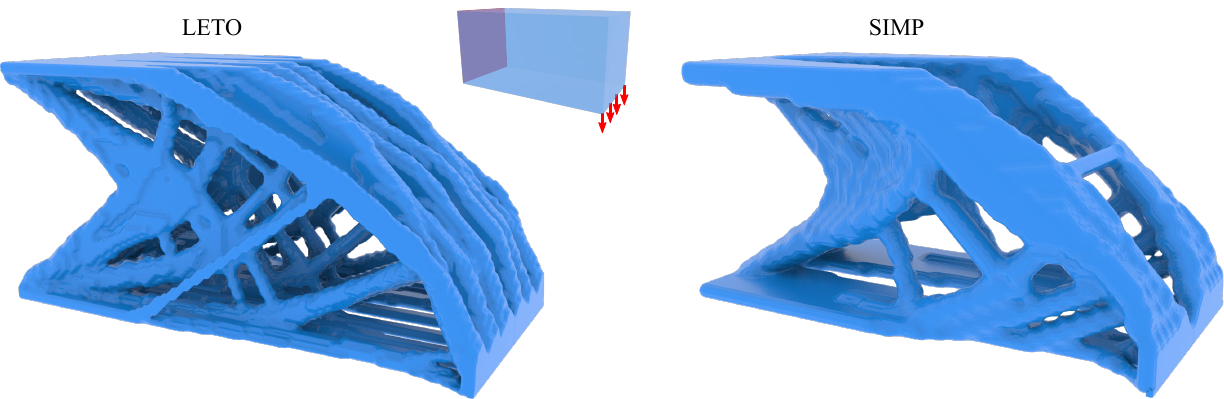}
    \caption{\textbf{3D beam.} 
    The results of SIMP and the proposed method \LETO for this 3D beam example is evaluated. 
    One ending plane of this cubic domain is fixed and forces are applied at the opposite plane's bottom edge.
    The compliance value evaluated at double-resolution grid for \LETO and SIMP is $3.334 \times 10^{-4}$, $3.527 \times 10^{-4}$ respectively.
    Richer thin supporting fibers can be seen in the result from the \LETO.
    }
    \label{fig:beam3d}
\end{figure}
Here \LETO is further evaluated on a 3D beam problem; see \cref{fig:beam3d}. In this example, the design domain is a cuboid of $1.6m \times 0.8m \times 0.8m$. One end of this cubic domain is fixed and a total force of $0.825N$ evenly distributes on the bottom edge of the opposite end. A symmetric boundary condition is utilized to reduce the simulation domain to half to save computational cost. The actual simulation grid resolution is $64\times32\times16$ with a spacing of $0.025m$. The target volume prescription is 20\%. With \LETO, the result forms a set of thin supporting structures. The compliance for SIMP evaluated at double resolution, \LETO evaluated at optimization resolution, and double resolution are $3.527 \times 10^{-4}$, $3.323 \times 10^{-4}$, $3.334 \times 10^{-4}$ respectively. The final volume reached by SIMP and \LETO is 20.0\% and 19.6\%. 

\subsubsection{3D Bridge}
\begin{figure}[ht]
    \centering
    \includegraphics[width=0.9\textwidth]{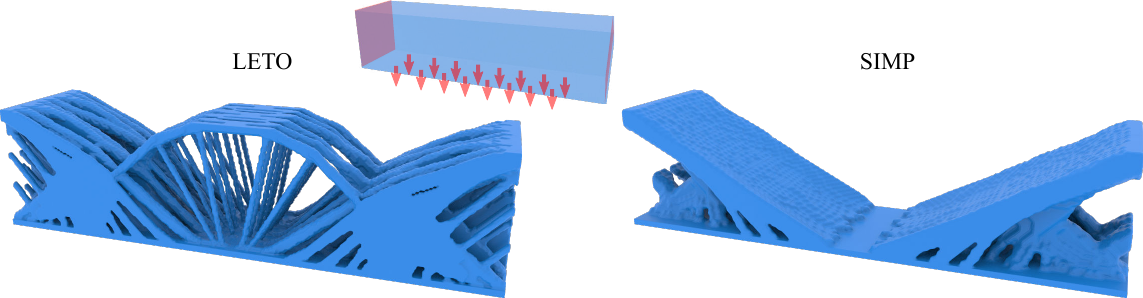}
    \caption{\textbf{3D bridge.} 
    A 3D brigade design is considered in this example, where the two ending planes of the cubic design domain is fixed, and a plane force is applied at the bottom.
    The compliance value evaluated at double-refined grid resolution for \LETO and SIMP is $1.643 \times 10^{-2}$, $1.892 \times 10^{-2}$ respectively.
    \LETO produces intricate supporting truss structures unseen in the results from the SIMP method.
    }
    \label{fig:bridge}
\end{figure}
In the second 3D example, \LETO is tested on a 3D bridge problem; see \cref{fig:bridge}. The design domain is a cuboid of a length of $4m$, a width of $1m$, and a height of $1m$. The two ending planes along the longest axis are fixed, and a plane force of a total $40.74N$ is added on the bottom (denoted by red arrows). Two symmetric boundary conditions are utilized to reduce the simulation domain to a quarter. The grid resolution in optimization is $160 \times 40 \times 40$ with a spacing of $0.025m$. The target volume is 20\%. The appearance of \LETO and SIMP appears to be significantly different in this example, where \LETO generates truss structures with rich fibers in the middle. The compliance value for SIMP evaluated at double resolution, \LETO evaluated at optimization resolution, and double resolution are $1.892 \times 10^{-2}$, $1.653 \times 10^{-2}$, $1.643 \times 10^{-2}$ respectively. The final volume reached by SIMP and \LETO at double resolution grid are 20.0\% and 19.4\%.

\subsubsection{Ablation Studies}

In this section, an ablation study on \LETO with different quadrature point distributions, together with a comparison between \LETO and SIMP with 1.2 filter radius are demonstrated.

First, two different variations of \LETO are compared with the proposed one:
\begin{itemize}
    \item \textbf{\LETO with Gaussian quadratures.} The quadratures within each cell are moved to Gaussian points. This means the sample points of the density field are not uniformly distributed as well.
    \item \textbf{Single-density \LETON.} The density field within each cell is a constant. The enforcement is achieved by setting each cell's density as the average density of the uniformly-distributed quadratures within that cell. When computing the density field, Gaussian quadratures are utilized to increase the numerical accuracy.
\end{itemize}

\begin{figure}[ht]
\centering
\includegraphics[width=0.9\textwidth]{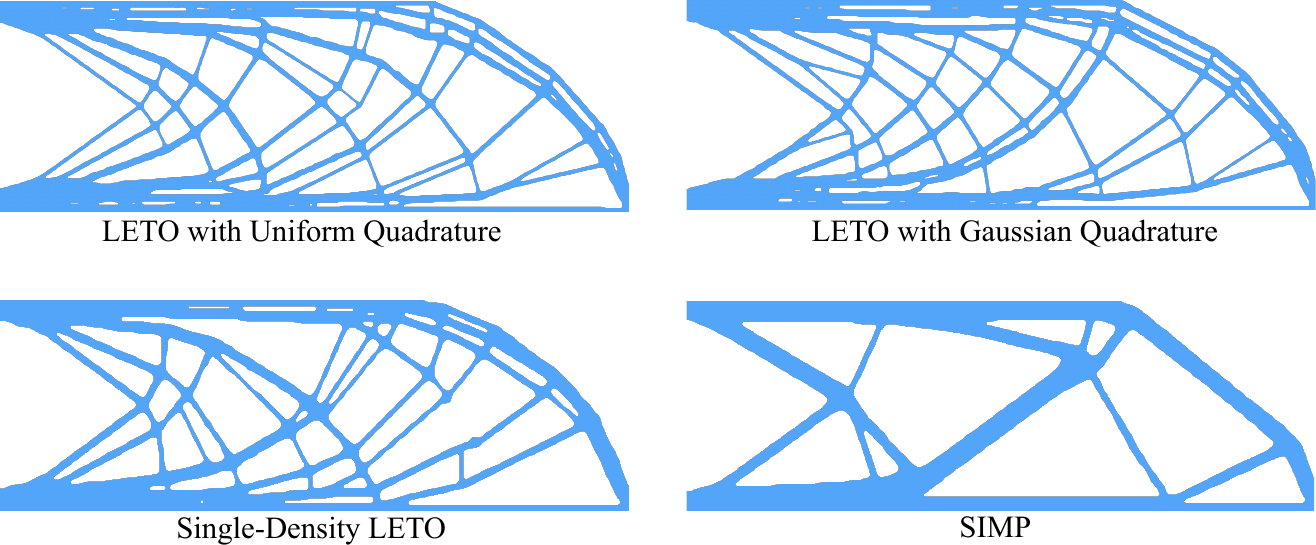}
\caption{Ablation study on variations of \LETO.}
\label{fig:quadrature_compare}
\end{figure}

The experiment setup up is the same as \cref{sec:concentrated_load_2d}. The optimization results are shown in \cref{fig:quadrature_compare}, where the top-left one is from the proposed \LETON. The top-right one and the bottom-left one are from the two variations of \LETON. The bottom right one is from SIMP. The quantitative comparison is included in \cref{tab:quadrature_compare}.

Among these versions of \LETON, the proposed one (\LETO with uniformly-distributed quadratures) performs the best. Compared to uniformly-distributed quadratures, Gaussian quadratures may cause bias on the sampling of the density field during the C2P transfer. The quantatitive comparison also shows the advantage of uniformly-distributed quadratures. The uni-density version of \LETO performs worst on this example. However, a notable distinction between the result from uni-density \LETO and the result from SIMP is that the result from uni-density \LETO has far more fine structures than SIMP. Likewise, the result from the proposed \LETO has more fine structures than uni-density \LETON. These comparisons show that the multi-density quadrature system and the particle-based material representation jointly contributes to the generated fine structures. 

\begin{table}[ht]
\centering
\begin{tabular}{ccccc}
\hline
Method & \begin{tabular}[c]{@{}c@{}}Quadrature\\ System\end{tabular} & \begin{tabular}[c]{@{}c@{}}Constant Density \\ in Cell\end{tabular} & \begin{tabular}[c]{@{}c@{}}Compliance\\ (refined grid)\end{tabular} & Volume \\ \hline
\LETO & Uniform & False & $1.241 \times 10^{-3}$ & 30.1\% \\
\LETO & Gaussian & False & $1.257\times 10^{-3}$ & 30.2\% \\
\LETO & Gaussian & True & $1.408\times 10^{-3}$ & 30.5\% \\
SIMP & Gaussian & True & $1.264\times 10^{-3}$ & 30.1\%\\
\hline
\end{tabular}
\caption{Quantative comparison between variations of the proposed \LETON.}
\label{tab:quadrature_compare}
\end{table}

\begin{figure}
    \centering
    \includegraphics[width=\textwidth]{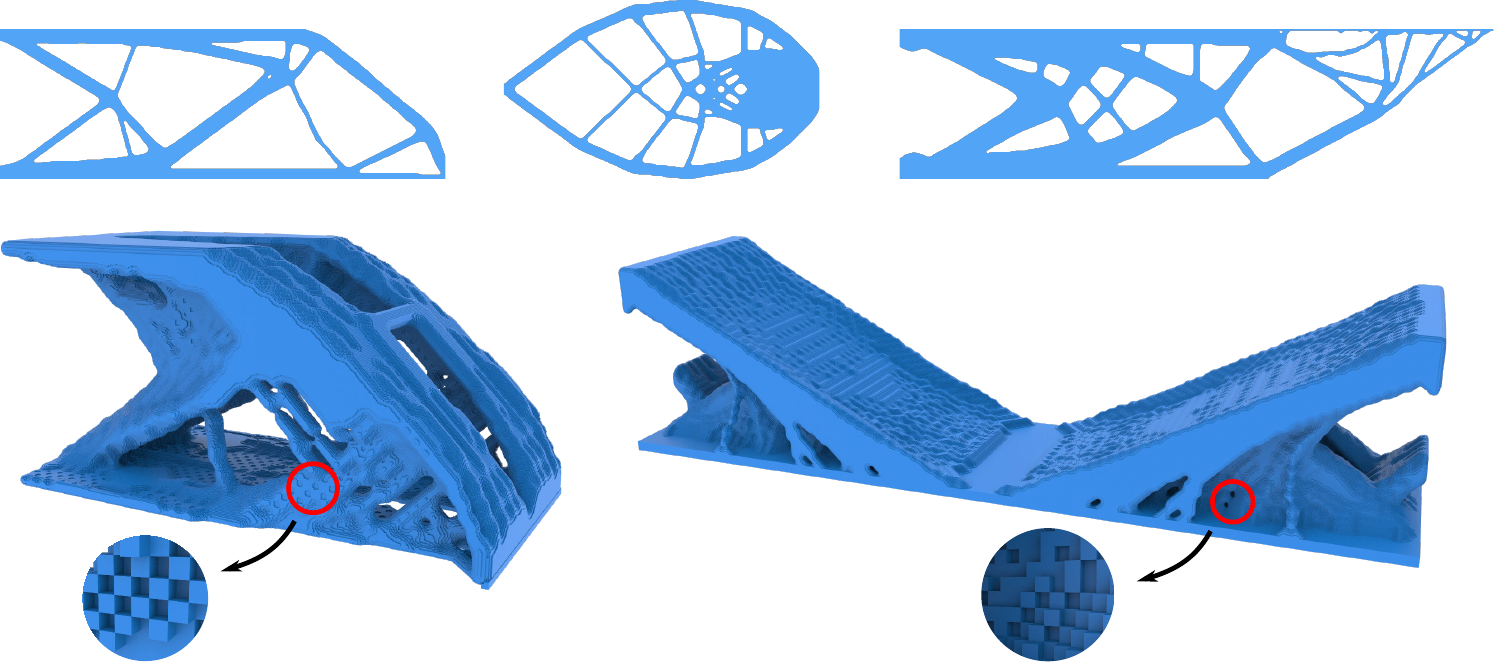}
    \caption{Linear topology optimization results of SIMP with filter radius 1.2.}
    \label{fig:simp-1-2-result}
\end{figure}

Then \LETO is also compared with SIMP with filter radius 1.2. In this case, the filter only touches the adjacent cells on three axes. The results of SIMP on the five linear topology optimization examples are shown in \cref{fig:simp-1-2-result}. The quantitative comparisons are shown in \cref{tab:simp-1-2-result}. \LETO still performs better in these examples and generate finer structures. In addition, 3D results from SIMP have checkerboard artifacts because of the small filter radius.

\begin{table}[ht]
\centering
\begin{tabular}{c|cc|cc}
\hline
\multirow{2}{*}{\begin{tabular}[c]{@{}c@{}}~~\\Experiment\end{tabular}} & \multicolumn{2}{c|}{LETO} & \multicolumn{2}{c}{SIMP (filter radius = 1.2)} \\ \cline{2-5} 
 & \begin{tabular}[c]{@{}c@{}}Compliance\\ (refined grid)\end{tabular} & Volume & \begin{tabular}[c]{@{}c@{}}Compliance\\ (refined grid)\end{tabular} & Volume \\ \hline
Concentrated-load beam & $1.243 \times 10^{-3}$ & 29.9\% & $1.316 \times 10^{-3}$ & 30.0\% \\
Michell truss & $5.054 \times 10^{-2}$ & 20.1\% & $5.055 \times 10^{-2}$ & 20.1\% \\
Distributed-load beam & $8.386 \times 10^{-3}$ & 39.7\% & $1.458\times 10^{-2}$ & 40.0\% \\
3D beam & $3.334 \times 10^{-4}$ & 19.6\% & $3.666 \times 10^{-4}$ & 20.0\% \\
3D bridge & $1.643 \times 10^{-2}$ & 19.4\% & $1.938\times 10^{-2}$ & 20.0\% \\ \hline
\end{tabular}
\caption{Quantitative comparison between \LETO and SIMP with filter radius 1.2.}
\label{tab:simp-1-2-result}
\end{table}

\subsection{Nonlinear Topology Optimization}
In this section, the robustness of LETO's nonlinear static equilibrium solver is illustrated by varying the force magnitude in large ranges. 
Two different objectives are also compared: the elastic potential energy and mean compliance $u^T f^{ext}$, where $f^{ext}$ is the nodal external force field and $u$ is the nodal displacement field. 
Many topology optimization methods for nonlinear elasticity only consider cases with small strains, thus utilize the mean compliance as the objective function, essentially a linearization of the elastic potential \cite{buhl2000stiffness,bruns1998topology,gea2001topology}. 
The two objectives are equivalent in linear elasticity up to a factor. 
The following examples show that they differ significantly under large force magnitude. 
Minimizing the mean compliance is equivalent to minimizing the displacements at force loading points. However, that does not necessarily minimize the elastic energy stored in the material. 
Buckled structures will appear to reduce potential energy when force magnitude is large. The examples shown are all optimized with nonlinear \LETO. 
The compliance and mean compliance value reported are all evaluated on a double refined grid as in the linear topology optimization examples.

\subsubsection{2D Long Beam}
\begin{figure}[ht]
    \centering
    \includegraphics[width=\textwidth]{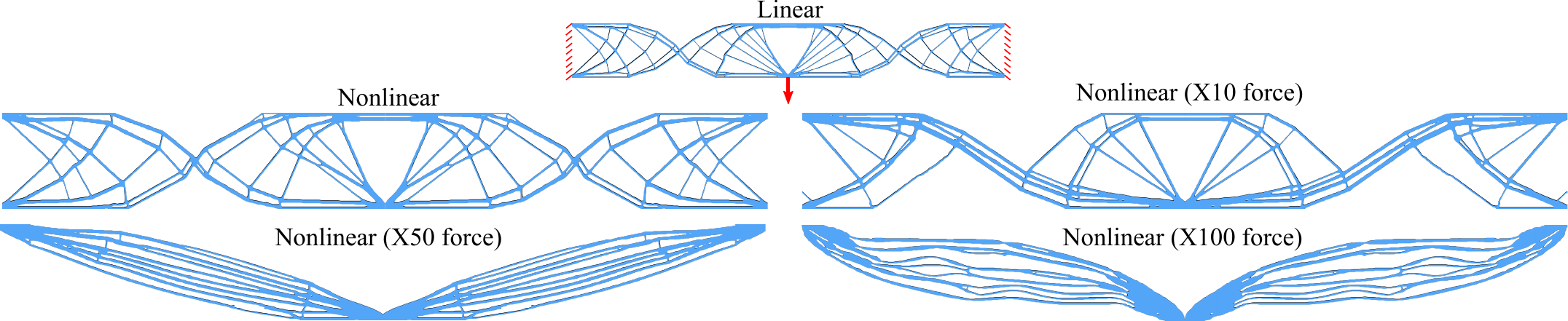}
    \caption{\textbf{2D long beam.} 
    The results of the proposed method under different force magnitudes are shown in this figure.
    The resulting material distribution differs sharply and a clear buckling behavior manifested.
    It can be seen that the proposed method is robust under large deformations.
    }
    \label{fig:long-beam}
\end{figure}

In this example, a concentrated force is applied at the bottom center of a long beam; see \cref{fig:long-beam}. The $4m \times 1m$ design domain is discretized with a grid resolution $800 \times 100$. The target volume is $20\%$. Force magnitude at $1N, 10N, 50N, 100N$ are tested. As shown in  \cref{fig:long-beam}, when the force magnitude is small, the result is close to that of linear topology optimization. As the force becomes larger, more and more fibers form between the force port and the two top corners. And finally, a lot of buckled fibers appear.
\par
\begin{figure}[ht]
    \centering
    \includegraphics[width=\textwidth]{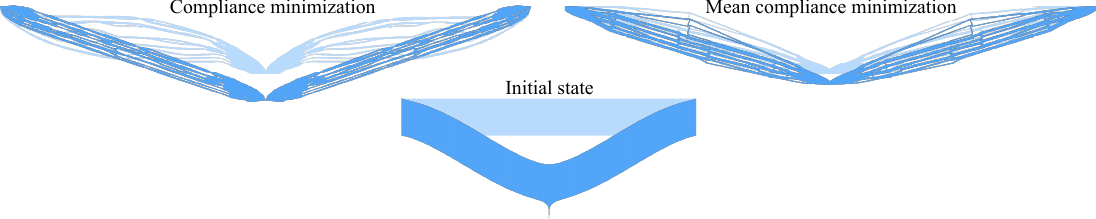}
    \caption{\textbf{2D long beam structure with different objectives.}
    The resulting material distribution of minimizing compliance and mean compliance under X100 force is shown in this figure.
    }
    \label{fig:long-beam_deformation}
\end{figure}

Under the maximal tested force magnitude, results of compliance (elastic potential energy) minimization and mean compliance minimization are compared; see \cref{fig:long-beam_deformation}. The undeformed and deformed states are differentiated by translucent and solid coloring. 
The final compliance of the result by minimizing elastic potential is $3.473$ and the mean compliance of the result by minimizing mean compliance is $3.921$. 
Although the displacement of force port is much smaller when minimizing mean compliance, the elastic energy stored in the structure is much larger: the compliance of the result by minimizing mean compliance is 8.535, which is significantly larger than the compliance value resulting from minimizing the elastic potential.

\subsubsection{3D Wheel}
\begin{figure}[ht]
    \centering
    \includegraphics[width=0.65\textwidth]{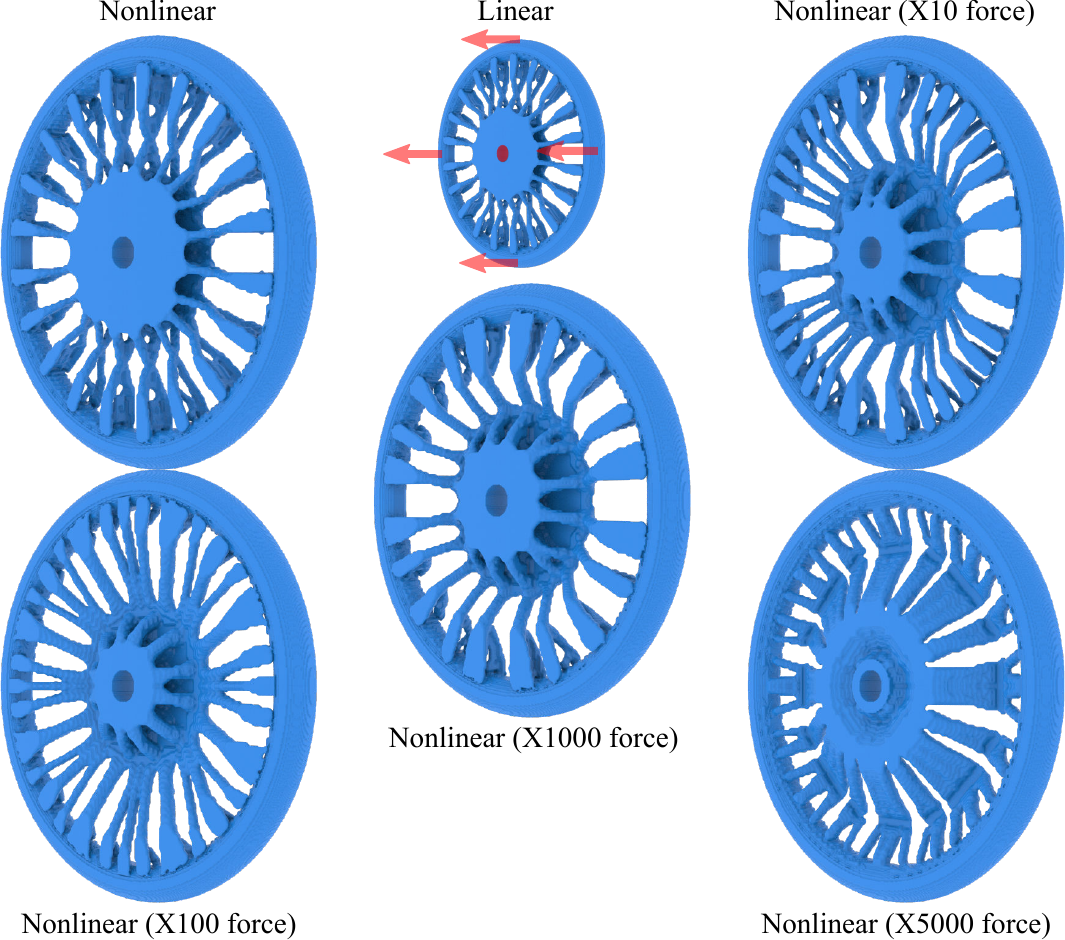}
    \caption{\textbf{3D wheel.} 
    In this example, a large range of force magnitudes is tested to further examine the robustness of the proposed method in 3D.
    Out-of-plane forces are applied to the wheel and the central region is fixed. }
    \label{fig:wheel}
\end{figure}

In this example, \LETO is tested on a 3D wheel design problem under normal forces; see \cref{fig:wheel}. The design domain is a flat cylinder bounded by a torus. The outer radius and inner radius of the torus are $1.2m$ and $1m$, respectively. The grid resolution is  $96 \times 96 \times 8$ with spacing $0.025m$. A small cylinder through the center of the wheel is fixed, and the outer torus is set to maintain solid. Forces are perpendicular to the wheel plane and are evenly exerted on a thin layer of the wheel's outer-most boundary. The magnitude of total force at $1.09\times 10^{-2}N, 1.09\times 10^{-1}N, 1.09N, 1.09\times 10^{1}N, 5.44\times 10^{1}N$ are tested. The target volume is $20\%$. When the force magnitude is small, the result of nonlinear elasticity is almost identical to linear elasticity, which is symmetric w.r.t the wheel plane. As the force magnitude becomes larger, the symmetry disappears, and the spokes become denser.
\par
\begin{figure}[ht]
    \centering
    \includegraphics[width=0.7\textwidth]{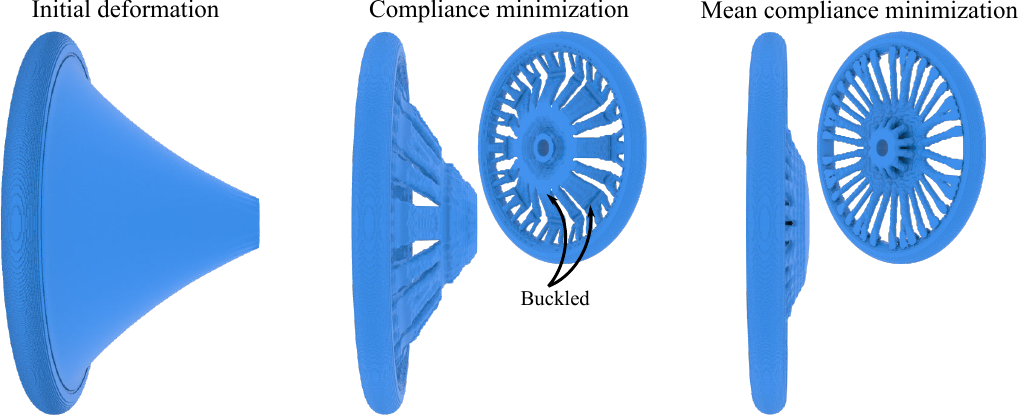}
    \caption{\textbf{3D wheel structure with different objectives.} 
    The difference in considering compliance and mean compliance as the objective function is further compared in this experiment, where a force is enforced to create significantly large initial deformation.
    Minimizing compliance objective delivers strong buckling behavior in the resulting structure.
    }
    \label{fig:wheel_deformed}
\end{figure}

Same as the 2D example, when the force is very large, buckled fibers appear. As shown in \cref{fig:wheel_deformed}, the buckled spokes open up in the deformed state. Result optimized by minimizing mean compliance is shown as well to compare. Although the overall deformation of it appears to be more moderate than the compliance-minimized result, it actually stores significantly more energy. The mean compliance of the mean-compliance-minimized result is $2.946$. However, the energy it contains is $6.960$, while the compliance of the compliance-minimized result is only $2.063$.

\section{Conclusion and Future work}
A new hybrid Lagrangian-Eulerian topology optimization method is proposed. MPM discretization is used to enable sub-cell resolution on the fly. The method produces intricate results with comparable and sometimes lower compliance at similar simulation costs than Eulerian methods. With a unified treatment, the proposed method optimizes the elastic potential as the compliance objective for both linear and highly nonlinear (\eg neo-Hookean) hyperelastic materials. Notably, the method robustly captures large deformation and buckling behaviors in nonlinear cases. 

The finite strain formulation in MPM allows the construction of a unified framework for general hyperelastic materials. With the ability to resolve sub-cell features, \LETO can be further extended in future work to optimize anisotropic, heterogeneous, and multi-scale materials. It would also be interesting to apply this framework to optimize different objectives, \eg, compliant mechanisms and task-oriented objectives for designing soft robots.

In the 2D long beam example for nonlinear elasticity, slight overlaps between some fibers happen under the maximal tested load when the equilibrium is solved with the narrow-band filter turned on. However, when the equilibrium is solved with weak material like SIMP, the non-invertible neo-Hookean constitutive model will push fibers apart. This brings the inspiration that non-invertible weak material could be a contact handling model. On the other hand, it is tricky to tune weak material's Young's modulus --- too large Young's modulus can provide non-realistic supporting force. On the other hand, too small Young's modulus may cause numerical issues, not to mention that the weak material wastes a lot of computational power. More robust handling of contact should be developed to enable contact-aware topology optimization.

Last but not least, the method relies on MMA. As mentioned, MMA requires careful parameter tuning to perform well. Even with the general and consistent regularizations for all examples on step length and the relative scaling between constraints and objective, it is still unclear whether the optimization parameters chosen in this paper are optimal. It is also unclear whether the current MMA parameters can conveniently adapt for extensions to more complex materials. Therefore, it would be meaningful to develop a more general and easy-to-setup optimizer, likely taking advantage of second-order design sensitivity information.

\bibliographystyle{unsrt}
\bibliography{references}  

\end{document}